\documentclass[twocolumn,english,prd,superscriptaddress,nofootinbib,floatfix,showpacs]{revtex4-1}
\usepackage{float,graphicx}

\usepackage[latin1]{inputenc}
\usepackage[T1]{fontenc}
\usepackage{amsmath}
\usepackage{amssymb}
\usepackage{pstricks}
\usepackage{graphicx}

\begin{document}

\title{Chameleons with Field Dependent Couplings}

\author{Philippe Brax}
\affiliation{ Institut de Physique Th\'eorique, CEA, IPhT, CNRS, URA 2306,
  F-91191Gif/Yvette Cedex, France.}
\author{Carsten van de Bruck}
\affiliation{ Department of Applied Mathematics, The University of Sheffield,
 Hounsfield Road, Sheffield S3 7RH, United Kingdom}
\author{David F. Mota}
\affiliation{Institute of Theoretical Astrophysics, University of Oslo, 0315 Oslo, Norway}
\author{Nelson J. Nunes}
\affiliation{Institut f\"ur Theoretische Physik,
Universit\"at Heidelberg, Philosophenweg 16,  D-69120 Heidelberg, Germany}
\author{Hans A. Winther}
\affiliation{ Institute of Theoretical Astrophysics, University of Oslo, 0315 Oslo, Norway}

\begin{abstract}
Certain scalar-tensor theories exhibit the so-called chameleon mechanism, whereby observational signatures of scalar fields are hidden by a combination of self-interactions and interactions with ambient matter. Not all scalar-tensor theories exhibit such a chameleon mechanism, which has  been originally found in models with inverse power run-away potentials and field independent couplings to matter. In this paper we investigate field-theories with field-dependent couplings and a power-law potential for the scalar field. We show that the theory indeed is a chameleon field theory. We find the thin-shell solution for a spherical body and investigate the consequences for E\"ot-Wash experiments, fifth-force searches and Casimir force experiments. Requiring that the scalar-field evades gravitational tests, we find that the coupling is sensitive to a mass-scale which is of order of the Hubble scale today.
\end{abstract}


\maketitle

%\begin{document}
\section{Introduction}
Modern cosmological observations strongly suggest that visible matter contributes only a few percent to the total energy budget. The rest is made of dark matter and dark energy. While dark matter is very
well motivated within particle physics, dark energy is harder to explain within particle physics models.
Scalar fields are natural candidates for dark energy, but the dark energy scalar field must be very light
to explain the accelerated expansion. In addition, its coupling to matter should be very small. Alternatively,
dark energy (and dark matter) might signal a breakdown of General Relativity on large scales.

In the last three decades, scalar fields have played an important role in both cosmology and particle physics (see e.g.
\cite{Linde,Binetruy} and references therein). The best motivated particle physics candidate for a scalar
field is the Higgs boson, part of the standard model of particle physics, which itself has yet to be observed.
Even though no scalar fields have ever been observed directly yet, they are a general feature of  high
energy physics beyond the standard model and are often related to the presence of extra-dimensions.
Scalar fields have been postulated as means to explain the early and late time acceleration of the
Universe. However, it is almost always the case that such fields interact with matter: either due to a direct
Lagrangian coupling or indirectly through a coupling to the Ricci scalar or as the result of quantum loop corrections. Both for inflation in the early universe and for dark energy, such couplings can lead to problems. In inflation, for example, couplings might destroy the flatness of the potential needed to drive a period of inflation. If there are scalar fields which permeate the universe today and have non-zero couplings to matter, then they would induce
an additional force in nature. If the scalar field self-interactions are negligible, then the experimental bounds on such a field are very strong: either the couplings to matter are much smaller than gravity, or the scalar fields are very heavy, so that they mediate a  short-ranged interaction.

However, a certain class of theories have been proposed, in which the scalar field(s) properties depend on the environment: these are the class of chameleon field theories, proposed by Khoury and Weltman \cite{Khoury:2003rn}, that employs a
combination of self-interaction and couplings to matter of the scalar-field to avoid the most restrictive of the current bounds.
In the models that they proposed, which from now on will be referred to as the standard chameleon model (SCM), a scalar field
couples to matter with gravitational strength, in harmony with general expectations from string theory, whilst, at the
same time, remaining relatively light on cosmological scales. It was found that local gravity constraints are (roughly) satisfied as long as the mass-scale of the potential satisfies $M\lesssim (1mm)^{-1}$. This coincides with the scale associated with the late time acceleration of the universe, and it is surprising that it should come from local experiments. We will, in this paper, show that this result carries over to other classes of chameleon models, (it will result from the current accuracy of the Eöt-Wash and Casimir experiments), in which the coupling becomes field dependent and hence is environment-dependent.

The chameleon with a constant coupling has been subject to many studies \cite{Brax:2004qh}-\cite{Tsujikawa:2009yf} to mention some. And most relevant experimental bounds have been calculated for the two fiducial potentials introduced by Khoury and Weltman. There have been very few studies on the different types of couplings\footnote{See \cite{Brax:2004ym} for a brief note on the power-law coupling $\beta(\phi) = \left(\frac{\lambda\phi}{M_{\rm pl}}\right)^n$.}. However, it would be important to investigate whether the chameleon mechanism is present in more general classes of models. Here, we will go one step further and generalise the chameleon mechanism to an inverse power-law coupling. In doing so the coupling to matter becomes dynamical and will be much smaller on earth than in the vacuum of space. When objects become big (in density and size) in a way defined later we will also have an additional suppression of the fifth-force by a thin-shell effect. We will show that our model  does not yield an explicit geometrical thin-shell, as found by solving the field-equation in the SCM. Instead we find an equivalent thin-shell factor which determines how much the resulting fifth-force is suppressed. Adopting the thin-shell name for our solution, we derive the expressions of the far-away field of thin-shelled bodies and show that it is independent of the parameters determining the coupling to matter. This is the same effect as found in \cite{Mota:2006fz} for the SCM.

This paper is divided into three main parts: in section II we study the behaviour of the scalar field inside and outside a spherical body. We find that the theory exhibits the chameleon mechanism and find the thin-shell solution. This allows us to make predictions for the chameleon behaviour  on earth and in the solar system. In section III we derive the expressions for the chameleon force law between different objects and ranges, which can be succinctly stated by introducing an effective coupling. In section IV we calculate the bounds on our parameters  from  the Eöt-Wash experiment, fifth-force searches, post-newtonian corrections and Casimir experiments.

We will show that the model allows for a very large local matter coupling, $|\beta,_{\phi_c}|M_{\rm pl}$, to be
compatible with all the available data. This is entirely due to the thin-shell effect. We will also show how non-linear effects ensure that the field value taken by the chameleon far away from a body with a thin-shell is independent of $\lambda$, the parameter that describes the strength of the coupling in the Lagrangian.
\subsection{Notation and conventions}
We will always work in units of $c\equiv 1$ and $\hbar\equiv 1$, the metric has the signature $(-,+,+,+)$ and we will use the convention $M_{\rm pl} \equiv \frac{1}{\sqrt{8\pi G}}$ for the Planck-mass. The frame referring to $g$ will be called the Einstein frame, and the frame referring to $\tilde{g}$ the Jordan frame. When speaking about the chameleon mass $m_{\phi}^2\equiv V_{\text{eff},\phi\phi}$ we refer to the mass of oscillations about a minimum of the effective potential. In looking at the field inside and outside a body the quantities of that body are referred to with a subscript $_c$ and the background with a subscript $_b$. For example the minimum of the effective potential inside (outside) a body is denoted by $\phi_c$ ($\phi_b$). When speaking about quantities such as $\beta,_{\phi}(\phi_b)$ we will sometimes simply  write $\beta,_{\phi_b}$.
\subsection{The Chameleon Action}
The action governing the dynamics of a general scalar-tensor theory is given by
\begin{equation}\label{cham_action}
S = \int dx^4 \sqrt{-g}\left[\frac{RM_{\rm pl}^2}{2} - \frac{1}{2}(\partial\phi)^2 - V(\phi) -\mathcal{L}_m(\tilde{g}_{\mu\nu},\psi_i)\right]
\end{equation}
where $g$ is the determinant of the metric $g_{\mu\nu}$, $R$ is the Ricci-scalar and $\psi_i$ are the different matter fields. The matter fields couple to $\tilde{g}_{\mu\nu}$ which is related to $g_{\mu\nu}$ via a conformal rescaling of the form
\begin{equation}\label{conformal_transform}
\tilde{g}_{\mu\nu} = A(\phi)^2g_{\mu\nu}
\end{equation}
The SCM corresponds to the choice $A(\phi)= e^{\frac{\beta\phi}{M_{\rm pl}}}$ where $\beta$ is a constant together with a run-away potential like $V(\phi) = M^4\left(\frac{M}{\phi}\right)^n$. Cosmological and local gravity experiments impose $\frac{\beta\phi}{M_{\rm pl}} \ll 1$ at least since the time of Big Bang Nucleosynthesis (BBN) so that in most applications of this model we can without loss of generality set $A = 1+\frac{\beta\phi}{M_{\rm pl}}$. This model has been found to be in agreement with experiments even for $\beta \gg 1$ provided one imposes a fine-tuning in the potential no worse than a cosmological constant. This is different from a minimally coupled scalar field for which fifth-force and equivalence principle experiments require a coupling strength much smaller than unity. In this work, we will study an inverse power coupling
\begin{equation}\label{coupling}
\log A(\phi)\equiv \beta(\phi)=\left(\lambda\frac{M_{\beta}}{\phi}\right)^k
\end{equation}
where $M_{\beta}$ is a mass-scale and $\lambda$ a dimensionless parameter. We will refer to this model as a chameleon due to similarities with the SCM, even though we do not know a priori whether  this model will produce a chameleon thin-shell suppression effect.
\subsection{The Chameleon Potential}
The most important ingredient in a chameleon field theory is that the effective potential has a minimum which depends on the local matter density. The simplest type of potential, for our coupling \eqref{coupling}, having this property is the power-law potential
\begin{equation}\label{power_potential}
V(\phi) = \sigma M^4\left(\frac{\phi}{M}\right)^n
\end{equation}
where $M$ is a mass scale, $n>0$, and $\sigma$ a dimensionless parameter. When $n\not=4$ we can fix $\sigma=1$ by redefining the mass-scale $M$. This potential gives rise to an effective potential, defined below, of the same type as in the SCM. Here $M$ can be any mass-scale, but in order for the chameleon to act as a dark-energy candidate we need $V(\phi_{\rm today})\sim \Lambda$ together with an equation of state $\omega \approx -1$. It is therefore convenient to set $M=M_{DE}=\Lambda^{\frac{1}{4}}$ and have the cosmological constant as part of the potential. In this case we can think of the potential as a Taylor expansion of a more complicated potential such as  $V=M^4\exp(\phi^n/M^n)$,  for $\phi \ll M$.

\subsection{The Field equation}
Variation of the action ($\ref{cham_action}$) with respect to $\phi$ yields the field-equation
\begin{equation}\label{full_field_equation}
\square\phi = V_{,\phi} + \sum_i \frac{2}{\sqrt{g}}\frac{\partial \mathcal{L}_m}{\partial g^{(i)}_{\mu\nu}}g^{(i)}_{\mu\nu}\beta_{,\phi}^{(i)}
\end{equation}
where the sum is over the different matter-species and we have allowed for different couplings to different species. Assuming that the matter fields $\psi^i$ do not interact with each other, each energy-momentum tensor (suppressing the $(i)$ for now)
\begin{equation}
\tilde{T}^{\mu\nu} = -\frac{2}{\sqrt{\tilde{g}}}\frac{\partial\mathcal{L}_m}{\partial \tilde{g}_{\mu\nu}}
\end{equation}
is conserved in the Jordan-frame \cite{Waterhouse:2006wv}
\begin{equation}\label{emjframe}
\tilde{\nabla}_{\nu}\tilde{T}^{\mu\nu} = 0
\end{equation}
where $\tilde{\nabla}$ is the Levi-Civita connection corresponding to the metric $\tilde{g}$. In the perfect fluid approximation where each matter species behaves as a perfect isentropic fluid with equation of state $\tilde{p}=\omega_i\tilde{\rho}$ we have
\begin{equation}
\tilde{T}^{\mu\nu}\tilde{g}_{\mu\nu}=-\tilde{\rho}+3\tilde{p} = -\tilde{\rho}(1-3\omega_i)
\end{equation}
Going to the Einstein frame we choose, without loss of generality, a Friedmann-Lemaître-Robertson-Walker (FLRW) background metric. The energy density $\rho$ in the Einstein-frame is the one that obeys the usual continuity equation $\rho \propto a^{-3(1+\omega_i)}$. Computing the Christoffel-symbol
\begin{equation}\label{csymb}
\tilde{\Gamma}^{\mu}_{\alpha\nu} = \Gamma^{\mu}_{\alpha\nu}+\frac{d\ln A}{d\phi}\left(\delta^{\mu}_{\alpha}\phi_{,\nu}+\delta^{\mu}_{\nu}\phi_{,\alpha}-g_{\alpha\nu}\phi^{,\mu}\right)
\end{equation}
and using $(\ref{emjframe})$ we find
\begin{equation}
\frac{d}{dt}\left(A_i^{3(1+\omega_i)}(\phi)\tilde{\rho}_ia^{3(1+3\omega_i)}\right)=0
\end{equation}
Thus
\begin{equation}
\rho_i = A_i^{3(1+\omega_i)}(\phi)\tilde{\rho}_i
\end{equation}
is the Einstein-frame density. With this definition, the equation of motion, in the Einstein-frame, reads
\begin{equation}
\square\phi = V_{,\phi} + \sum_i\rho_i(1-3\omega_i)A_{i,\phi}A_i^{(1-3\omega_i)}
\end{equation}
and we see that the field equation for $\phi$ can be written $\square\phi = V_{\rm eff,\phi}$, where the effective potential is given by
\begin{equation}
V_{\text{eff}}(\phi) = V(\phi) + \sum_i\rho_i A_i^{(1-3\omega_i)}(\phi)
\end{equation}
To simplify things we will assume that all the different matter species couple to $\phi$ with the same $A(\phi)$ and we will only consider non-relativistic matter where $\omega_i\approx 0$. With these assumptions, the effective potential reduces to
\begin{align}
V_{\text{eff}}(\phi) = V(\phi) +  \rho A(\phi)
\end{align}
Note that since the matter fields couple to $\tilde{g}$, the geodesics of a test-particle will be the geodesics of this metric, and $\tilde{\rho}$ is the physical density. We do not need to be too careful about this since, as we will show, in all practical applications we will have $A(\phi)\approx 1$ and the two densities are essentially the same.
\subsection{Minimum of the effective potential}
The minimum of the effective potential is determined by the equation $V_{\text{eff},\phi}=0$ which gives
\begin{equation}
\phi_{\text{min}} = M\left(\frac{\lambda M_{\beta}}{M}\right)^{\frac{k}{n+k}}\left(\frac{k\rho}{\sigma nM^4}\right)^{\frac{1}{n+k}}
\end{equation}
The chameleon mass at the minimum is given by
\begin{align}\label{cham_mass}
\begin{array}{c}m_{\phi}^2 \equiv V_{\text{eff},\phi\phi}(\phi_{\text{min}}) = \frac{k(n+k)\rho}{\lambda^2M_{\beta}^2}\left(\frac{\lambda M_{\beta}}{\phi_{\text{min}}}\right)^{k+2}\\
=M^2k(n+k)(\sigma n/k)^{\frac{k+2}{n+k}}\left(\frac{\rho}{M^4}\right)^{\frac{n-2}{n+k}}\left(\frac{\lambda M_{\beta}}{M}\right)^{\frac{k(n-2)}{n+k}}\end{array}
\end{align}
where we have used that in contrast to the standard chameleon where $m_{\phi}^2=V,_{\phi\phi}$ we now have to take in account the contribution from the term $\beta,_{\phi\phi}\rho$. But we can ignore the term $\rho\beta^2,_{\phi}$ which is valid as long as $\beta(\phi)\ll 1$. From (\ref{cham_mass}) we see that the field can only be a chameleon for $n>2$.
\subsection{An equivalent formulation}
We redefine the field by introducing $\chi = \frac{M_{\beta}}{\phi} M_{\rm pl}$. The coupling yields
\begin{align}
	\beta(\chi) = \left(\frac{\lambda\chi}{M_{\rm pl}}\right)^k
\end{align}
which becomes that of the SCM for $k=1$. Our power-law potential (\ref{power_potential}) becomes
\begin{align}
	V(\chi) = \sigma M^4\left(\frac{M_*}{\chi}\right)^n
\end{align}
where $M_* = \frac{M_{\beta}M_{\rm pl}}{M}$. When $M_{\beta}=\frac{M^2}{M_{\rm pl}}$ we have $M_*=M$ and the potential is identical to the Rattra-Peebles potential often used in the SCM. With this choice for $M_{\beta}$, the full action can be written
\begin{eqnarray}\label{equivalent_action}
	S &=& \int dx^4 \sqrt{-g}\left[\frac{RM_{\rm pl}^2}{2} - \frac{1}{2}\left(\frac{M}{\chi}\right)^4(\partial\chi)^2 - V(\chi) \right. \nonumber \\ &-& \left. \mathcal{L}_m\left(e^{2\left(\frac{\lambda\chi}{M_{\rm pl}}\right)^k}g_{\mu\nu},\psi_i\right)\right]
\end{eqnarray}
and we see that (for $k=1$) it is only the kinetic terms that distinguish our model from the SCM. The fine-tuning in the coupling sector is removed and we are left with only one fine-tuned mass-scale in the action.

The field equation is given by
\begin{align}
\begin{array}{c}
\square\chi - \frac{2}{\chi}(\nabla_{\mu}\chi)^2 = \left(\frac{\chi}{M}\right)^4V_{\text{eff},\chi}\\
V_{\text{eff}}(\chi) = M^4\left(\frac{M}{\chi}\right)^n + \rho e^{\left(\frac{\lambda\chi}{M_{\rm pl}}\right)^k}
\end{array}
\end{align}
which is significantly more complicated to work with than (\ref{full_field_equation}) so we will use the original formulation.
\subsection{The Coupling Scale}
In the background today, taking $\lambda=1$, we have
\begin{equation}
\frac{\phi_b}{M_{\beta}} \sim \left(\frac{M}{M_{\beta}}\right)^{\frac{n}{n+k}}\left(\frac{M_{DE}}{M}\right)^{\frac{4}{n+k}}.
\end{equation}
For the model to be in agreement with experiments we must require $\beta(\phi_b) \ll 1$. This constrains
\begin{equation}
M_{\beta}\ll M_{DE}\left(\frac{M}{M_{DE}}\right)^{\frac{n-4}{n}}
\end{equation}
showing the need to fine-tune $M_{\beta}$. We fix $M_{\beta}$ by the requirement that the equivalent action (\ref{equivalent_action}) is of the same form as the SCM when $M=M_{DE}\sim (1mm)^{-1}$. This fixes
\begin{align}
	M_{\beta} = \frac{M^2}{M_{\rm pl}}\sim H_0
\end{align}
This choice also ensures that the coupling $|\beta,_{\phi_c}|M_{\rm pl}$ of a $\rho_c\sim 1g/cm^3$ body is of order 1 when $M=M_{DE}$. The term $|\beta,_{\phi}|M_{\rm pl}$ plays the same role in this model as $\beta$ does in the SCM, but here this factor is field-dependent. In the remaining of this article we take $M_{\beta} = H_0$ so that
\begin{align}
\beta(\phi) = \left(\lambda\frac{H_0}{\phi}\right)^k
\end{align}
is our coupling.
\section{Spherical Solutions to the field equation}
The field equation in a static spherical symmetric metric with weak gravity reads
\begin{equation}
\ddot{\phi}+\frac{2}{r}\dot{\phi} = V,_{\phi} + \rho\beta,_{\phi}
\end{equation}
where we have assumed $\beta(\phi)\ll 1$. We study  solutions inside and outside a spherical body of constant density $\rho_c$ (e.g. the earth) in a background of a very low density $\rho_b\ll \rho_c$,
\begin{align}
	\rho = \left\{\begin{array}{c}\rho_c \text{ for } r<R\\ \rho_b \text{ for } r>R\end{array}\right.
\end{align}
and impose the boundary conditions
\begin{align}
\begin{array}{lcl}
	\left.\frac{d\phi}{dr}\right|_{r=0} &=& 0\\
	\left.\frac{d\phi}{dr}\right|_{r=\infty} &=& 0
\end{array}
\end{align}
The first condition follows from the spherical symmetry around $r=0$ and the second one implies that the field converges to the minimum of the effective potential, $\phi_b$, in the far-away background. If $m_c R \ll 1$, the chameleon acts approximately as a linear scalar field whereas in the case  $m_cR\gg 1$ the full non-linearity of the field equation comes into play.
\subsection{Case 1: the Thick-shell $m_cR\ll 1$}
\begin{figure}[htbp]
	\centering
		\includegraphics[width=0.9\columnwidth]{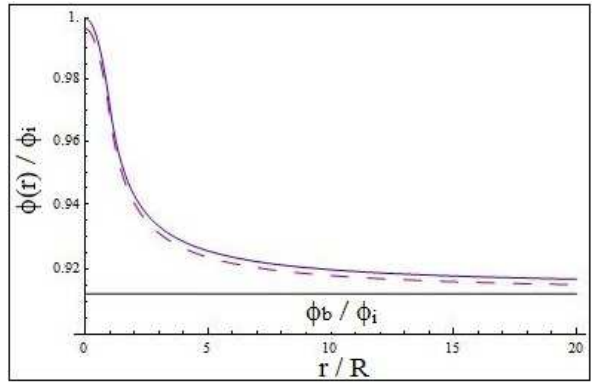}
	\caption{Numerical field profile for $(n,k)=(10,1)$ and $m_c R = 10^{-3}$ together with the analytical approximation (dashed line). The analytical approximation is seen to be a very good match to the actual solution.}
	\label{fig:linear_regime}
\end{figure}

In this case the initial value satisfies $\phi(0)\equiv \phi_i\ll\phi_c$ and the approximation $V_{\text{eff},\phi}\approx \beta,_{\phi}\rho_c$ is valid inside the body. Since this driving force is relatively small, we approximate $\beta_{,\phi}\rho_c \approx \beta_{,\phi_i}\rho_c$. Solving the field equation is now straightforward and the solution reads
\begin{equation}
\phi \approx \phi_i - \frac{|\beta,_{\phi_i}|\rho_cr^2}{6} \text{ for } 0<r<R
\end{equation}
where we have used absolute values since $\beta,_{\phi}<0$. This solution corresponds to the thick-shell solution in the SCM, which is not surprising since the non-linearities in the field equation are negligible. Outside the body we assume that the linear approximation is valid leading to a Yukawa profile $\phi = \phi_b +\frac{ARe^{-m_br}}{r}$. Matching the two solutions at $r=R$ leads to
\begin{align}
A = \frac{|\beta,_{\phi_i}|\rho_cR^2}{3}
\end{align}
with $\phi_i$ determined through
\begin{align}\label{phiieq}
\phi_i - \frac{|\beta,_{\phi_i}|\rho_cR^2}{2} = \phi_b
\end{align}
Defining $m_i^2 = \rho_c\beta,_{\phi\phi}(\phi_i)$, the chameleon mass at the centre of the object, this last expression can be rewritten as
\begin{align}\label{mieq}
(m_iR)^2 = 2(k+1)\left(1-\frac{\phi_b}{\phi_i}\right)
\end{align}
and  the chameleon takes a value at the centre of the body corresponding to a mass  $m_i \sim \frac{1}{R}$. This also shows that the approximation used inside the body is valid since the field  undergoes a $\frac{\phi(R)-\phi(0)}{\phi(0)} \lesssim \mathcal{O}(1)$ change. The initial value, $\phi_i$, can be rewritten in a  more compact fashion for $\phi_i\gg\phi_b$ as
\begin{align}\label{phii}
\phi_i = \phi_c\left(\frac{(m_c R)^2}{2(n+k)}\right)^{\frac{1}{k+2}}.
\end{align}
If $m_c R$ is really small we have $\phi_i \approx \phi_b$ and the field inside the body is just a small perturbation in the background.  To summarise, the solution is
\begin{align}
\begin{array}{lcll}
\phi &=& \phi_i - |\beta,_{\phi_i}|\frac{\rho_c r^2}{6} &\text{ for } 0<r<R\\
\phi &=& \phi_b + \frac{|\beta,_{\phi_i}|}{4\pi}\frac{M_1 e^{-m_br}}{r}  &\text{ for } R<r\\
\phi_i &=& \phi_b + \frac{|\beta,_{\phi_i}|\rho_cR^2}{2}.
\end{array}
\end{align}
The far-away  field is proportional to the coupling $\beta,_{\phi}$ evaluated inside the body (or equivalently at the surface), just like for the SCM. Let us mention that two bodies with $m_c R \ll 1$  attract each other with an attractive force of magnitude
\begin{align}
	\vert F_{\phi}\vert  = 2\beta,_{\phi_i^{(1)}}\beta,_{\phi_i^{(2)}}M_{\rm pl}^2\frac{GM_1M_2(1+m_br)e^{-m_br}}{r^2}.
\end{align}
The relative strength  to gravity can be read off as $2\beta,_{\phi_i^{(1)}}\beta,_{\phi_i^{(2)}}M_{\rm pl}^2$ which  is maximal  for bodies where $\phi_i\approx \phi_b$. If a body increases in size the strength of the fifth-force decreases. In contrast with the SCM, this suppression appears  even for bodies without a thin-shell ($m_c R\ll 1$). See Fig.~\ref{fig:linear_regime} for a plot of a $m_cR\ll 1$ profile compared to the analytical approximation found above.

Note that outside the body we have assumed that the Yukawa profile is a good approximation. When $\phi_i\gg\phi_b$ we have $\phi(R)\gg\phi_b$ meaning that the approximation $V_{\text{eff},\phi}\approx m_b^2(\phi-\phi_b)$ is not  valid right outside $r=R$. In these cases, the driving term $V,_{\phi}$ can be neglected relative to the friction term, leading to the same $1/r$-profile. This approximation is valid up to  the region where $m_b r \sim 1$ or equivalently $\phi\sim \phi_b$ which leads to the Yukawa solution used above. As $m_b R\ll 1$, we can add the exponential factor to the solution  outside $r=R$. The numerical results show that the analytical solutions found above match the actual solutions to a good level of accuracy.
\subsection{Case 2: the thin-shell $m_cR\gg 1$}
\begin{figure}[htbp]
	\centering
		\includegraphics[width=0.9\columnwidth]{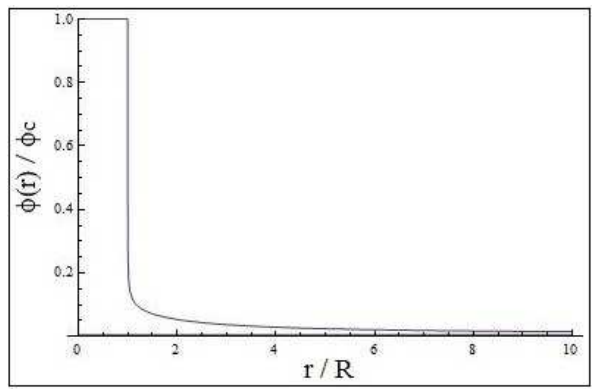}
	\caption{The Thin-shell profile for the earth when $(n,k)=(10,1)$ and $m_c R = 10^{6}$.}
	\label{fig:thinshellprofile}
\end{figure}

In this case the field starts out very close to the minimum inside the body: $\phi_i\approx \phi_c$. The field will remain close to the minimum throughout the body making the linear approximation $V_{\rm eff,\phi} = m_c^2(\phi-\phi_c)$ valid in $0<r<R$. This case is similar to the thin-shell solution in the SCM, with the exception that we do not have this explicit thin-shell. Nevertheless we will adopt the thin-shell name for our solutions. Right outside a thin-shelled body the approximation $V_{\text{eff},\phi}\approx V,_{\phi}$ is valid and we must solve
\begin{equation}\label{outapprox}
\ddot{\phi}+\frac{2}{r}\dot{\phi} \approx  n\sigma M^3\left(\frac{\phi}{M}\right)^{n-1} \text{ for } R<r<R^*
\end{equation}
where $R^*$ is the point where the coupling term, $\rho_b\beta,_{\phi}$, becomes relevant again. When $n=1$ or $n=2$ we can solve $(\ref{outapprox})$ as it stands. In these cases the field is not a chameleon since the chameleon mass increases as the density decreases. In the general case we will need certain approximations to find a solution.
\\\\
In Fig.~\ref{fig:thinshellprofile} we plot a thin-shelled solution for the earth in the cosmological background (density equal to the average cosmological density) for $n=10$.

Inside the body the field is  very close to the minimum and remains there throughout most of the body, except near the surface where the field undergoes a small change. Linearizing the effective potential around $\phi_c$: $V_{\text{eff},\phi}=m_c^2(\phi-\phi_c)$, we can find the solution close to $r=0$ that matches the initial condition
\begin{align}\label{interior_solution}
\begin{array}{lr}
	\phi = \phi_c\left(1-\tau\frac{\sinh(m_c r)}{m_c r}\right) &\text{ in } 0<r<R\\
	\tau = \frac{\phi_c-\phi(0)}{\phi_c}\ll 1.
\end{array}
\end{align}
The solution is valid as long as the linear term in the Taylor expansion of $V_{\text{eff},\phi}$ dominates over the higher order terms, which gives the condition
\begin{align}\label{linear_condition}
\begin{array}{lr}
	\left|\frac{\phi-\phi_c}{\phi_c}\right| < \frac{2}{|n-k-3|} &\text{ for } n-k-3\not= 0\\
	\left|\frac{\phi-\phi_c}{\phi_c}\right| < \left(\frac{6}{(k+1)(k+2)}\right)^{1/2} &\text{ for } n-k-3= 0.
\end{array}
\end{align}
The largest value of $|\phi-\phi_c|$ inside the body occurs at $r=R$ and we will later check that this value satisfies the condition above. Defining $\delta \equiv \tau \frac{\sinh(m_c R)}{m_c R}$ we have that the field value and derivative at $r=R$ satisfy
\begin{align}\label{phiR}
\begin{array}{ccl}
	\phi_R &=& (1-\delta)\phi_c\\
	\dot{\phi}_R &=& -\delta m_c\phi_c
\end{array}
\end{align}
Outside the body the potential is very steep, so the field  drops very quickly and  the friction term $\frac{2}{r}\dot{\phi}$ is, initially, negligible compared to the driving force $V,_{\phi}$, implying that
\begin{align}
	\ddot{\phi} \approx V,_{\phi}
\end{align}
To simplify the analysis we define $\psi \equiv \frac{\phi}{\phi_c}$, $x=\frac{r}{R}$  and $\frac{d}{dx} \equiv '$ so that we can write the equations in a dimensionless form as
\begin{align}\label{thinshell_approx}
	\psi'' = \frac{(m_cR)^2}{n+k}\psi^{n-1}
\end{align}
which has the solution
\begin{align}\label{thinshell_profile}
\begin{array}{ccl}
	\psi &=& \frac{\psi_R}{\left[1+a(r/R-1)\right]^{\frac{2}{n-2}}}\\
	a &=& \frac{m_c R(n-2)}{\sqrt{2n(n+k)}}(1-\delta)^{\frac{n}{2}-1}.
\end{array}
\end{align}
Matching to the solution for  $r<R$, using (\ref{phiR}), we find
\begin{align}\label{delta_eq}
	\frac{\delta^2}{(1-\delta)^n} = \frac{2}{n(n+k)}
\end{align}
which determines\footnote{The reader may question this result, since when $\delta$ is determined the full solution in $0<r<\infty$ is known (at least from a numerical point of view), but it is derived without considering the behaviour at large $r$ yet. What this result really states is that the solution that converges to $\phi_b$ in the background will have to correspond to a particular initial value of order $\delta$ at $r=R$. This result is confirmed by the numerics.} $\delta$. Numerically we find $\delta \approx 0.086$ when $(n,k)=(10,1)$ and $\delta \approx 0.200$ when $(n,k)=(4,1)$ for all $m_c R > 10$ in very good agreement with the formula above. If we now go back and put this value for $\delta$ into (\ref{linear_condition}) we see that the linear approximation is valid for all reasonable values of $(n,k)$. As an example, take $n=4$, upon using (\ref{delta_eq}) we find $\delta^2 \approx \frac{1}{2(k+4)}$ which in (\ref{linear_condition}) gives the condition
\begin{align}
	1 < \frac{8(k+4)}{(k-1)^2} ~~~\to ~~~ k<12.
\end{align}
As the field rolls down along the potential, it reaches  a point $r=R^*$ where the driving force satisfies\footnote{In the case where $m_bR\gg 1$ the field will have settled at the minimum before this happens. Since this corresponds to a very short ranged force we will not consider it here.}
\begin{align}
	F_{\rm driving} = \frac{(m_cR)^2}{n+k}\psi^{n-1} < 1
\end{align}
and from here on the dynamics of $\psi$ are determined by the friction term which we have neglected until now. The field equation reads
\begin{align}
	\psi''+\frac{2}{x}\psi' \approx 0 \text{ for } R^*<r
\end{align}
with the solution
\begin{align}\label{farfield_solution}
	\psi \approx \psi_b + \frac{A R^*}{r} \text{ for } R^*<r
\end{align}
for some $A$. This solution is  valid until we reach the region where the driving force has to be taken into account again. This is the case when $m_b r\sim 1$ or equivalently $\psi\sim \psi_b$ and alters the solution by adding a Yukawa exponential $e^{-m_br}$ to the $1/r$ term. Again since $m_b R^* < 1$ we can incorporate this by adding this term to (\ref{farfield_solution}) as
\begin{align}\label{thinshell_profile2}
	\psi \approx \psi_b + \frac{A R^*e^{-m_b(r-R^*)}}{r} \text{ for } R^*<r<\infty
\end{align}
The matching of (\ref{thinshell_profile}) and (\ref{thinshell_profile2}) at $r=R^*$, defining $\Delta = \frac{R^*-R}{R}$, implies the identifications
\begin{align}
\begin{array}{ccl}
	\psi_b + A &=& \frac{\psi_R}{(1+a\Delta)^{\frac{2}{n-2}}}\\
	A &=& \frac{\psi_R}{(1+a\Delta)^{\frac{2}{n-2}}}\frac{(1+\Delta)a}{1+a\Delta}\frac{2}{n-2}
\end{array}
\end{align}
When $m_b R < 1$ we  have $\psi(R^*)=A+\psi_b \gg \psi_b$ which leads to
\begin{align}\label{feq_solution}
\begin{array}{ccl}
	\Delta &=& \frac{2}{n-4}\\
	AR^* &=& \frac{B R}{(m_c R)^{\frac{2}{n-2}}} = \frac{B R}{(m_b R)^{\frac{2}{n-2}}}\psi_b\\
	B &=& \left(\frac{n(n+k)}{2}\right)^{\frac{1}{n-2}}\left(\frac{n-2}{n-4}\right)^{\frac{n-4}{n-2}}
\end{array}
\end{align}
where we have used $a\Delta \gg 1$ in order to simplify the solutions.  This derivation  does not apply for $n=4$. A similar derivation shows that (\ref{feq_solution}) is valid for $n=4$ when one takes the limit $n\to 4$ in the expression for $B$. Let us summarise the solutions we have found:
\begin{align}
\begin{array}{lll}
	\phi \approx& \phi_c &\text{ for }~~~~r<R\\
	\phi \approx& \frac{(1-\delta)\phi_c}{(1+a(r/R-1))^{\frac{2}{n-2}}} &\text{ for }~~~~R<r<R^*\\
	\phi \approx& \phi_b + \frac{\phi_cB}{(m_c R)^{\frac{2}{n-2}}}\frac{Re^{-m_b (r-R^*)}}{r} &\text{ for }~~~~R^*<r.
\end{array}
\end{align}
Defining the effective coupling in the thin-shell case via
\begin{align}\label{far_field_betaeff}
\phi = \phi_b + \frac{\beta_{\text{eff}}}{4\pi M_{\rm pl}}\frac{M_1e^{-m_br}}{r} ~~~~\text{ for }~~~~ R^*<r
\end{align}
we have that
\begin{align}\label{eq:betaeff}
	\beta_{\text{eff}} = \frac{4\pi M_{\rm pl}}{M_1}\left(M R\right)^{\frac{n-4}{n-2}}\left(\frac{n-2}{n-4}\right)^{\frac{n-4}{n-2}}\left(2\sigma\right)^{-\frac{1}{n-2}}
\end{align}
which is independent of the parameters defining the coupling. Thus, the exterior profile of a thin-shelled body depends only on the radii $R$, the mass $M_1$ and the potential parameters. In the SCM, using the same potential as we do, the same effect was found in \cite{Mota:2006fz}. Our expression for the effective coupling (\ref{eq:betaeff}) agrees with \cite[Eq. 24]{Mota:2006fz}. This similarity should not come as a surprise since a thin-shell solution is associated with dominating self-interactions.

\begin{figure}%
\centering
\includegraphics[width=0.9\columnwidth]{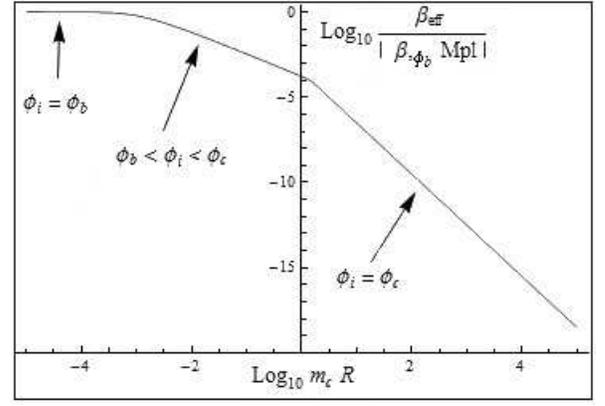}%
\caption{The effective coupling for a spherical body with constant density. When the body is very small the field inside the body is the same as the background, $\phi_b$, leading to a big coupling. Then as the radius gets bigger the field inside the body starts moving away from the background and the coupling decreases. Finally when we reach $m_c R > 1$, the field inside the body settles at $\phi_c$ and   develops a thin-shell such that the coupling starts to decrease like $1/R^3$.}
\label{fig:betaeff}%
\end{figure}

To justify the thin-shell name for our solutions, we define a thin-shell factor via
\begin{align}
\beta_{\text{eff}} = |\beta,_{\phi_c}|M_{\rm pl}	\frac{3\Delta R}{R}
\end{align}
and we find
\begin{align}
	\frac{\Delta R}{R} = \frac{\phi_c}{|\beta,_{\phi_c}|\rho_c R^2}\frac{B}{(m_c R)^{\frac{2}{n-2}}} \sim \frac{1}{(m_cR)^{\frac{2(n-1)}{n-2}}}
\end{align}
This factor determines how much of the mass of the body contributes to the fifth-force. As $m_c R \gg 1$, we have $\frac{\Delta R}{R} \ll 1$ and thus $\beta_{\text{eff}} \ll |\beta,_{\phi_c}|M_{\rm pl}$. If we extend this definition and set $\beta_{\text{eff}}=|\beta,_{\phi_i}|M_{\rm pl}$ when $m_c R\ll 1$ then (\ref{far_field_betaeff}) is valid for all bodies. See Fig.~\ref{fig:betaeff} for a plot of the effective coupling as a function of the radius of the body.
We note that for the special case when $n=4$ the far-away field can be written as
\begin{align}
	&\phi \approx \phi_b + \frac{\phi_b}{m_b} \frac{e^{-m_br}}{r}~~~~\text{ for }~~~~R^*<r
\end{align}
which is completely independent of the parameters $\rho_c$ and $R$ describing the body, and depends only on the background. Likewise $\beta_{\text{eff}}$ only depends on the mass of the body.
We are interested in the cases where the field has a large range in the solar-system together with thin-shelled planets $m_c R \gg 1$. From $\frac{m_c}{m_b} = \left(\frac{\phi_c}{\phi_b}\right)^{\frac{n-2}{2}} = \left(\frac{\rho_c}{\rho_b}\right)^{\frac{n-2}{2(n+k)}}$ we see that having a small $k$ and a large $n$ gives the largest ratio $\frac{m_c}{m_b} \sim \left(\frac{\rho_c}{\rho_b}\right)^{\frac{1}{2}}$. For the case of the earth $\rho_c\sim 1g/cm^3$ in a background of the average solar system density $\rho_b \sim 10^{-24}g/cm^3$ we find $\frac{m_c}{m_b} \sim 10^{12}$. It is therefore possible for the field to have a range as large as $m_b^{-1}\sim  10^{15}m \sim 10^4 Au$ and at the same time to have a thin-shelled earth $m_c R \gg 1$. This is the same as found in the SCM \cite{Khoury:2003rn}.

In the SCM, the coupling is easily identified as the parameter that multiplies $\phi$ in the matter-Lagrangian. Here we have a coupling that varies from place to place and is in general given by $\beta_{\text{eff}}$ defined above. A test particle in a region where $\phi\sim\phi_0$ will experience a coupling $|\beta,_{\phi_0}|M_{\rm pl}$. The coupling becomes smaller in a high-density environment and the largest value is achieved in the cosmological background. One can say that the chameleon effect in this model is twofold: first the coupling decreases as the environment gets denser and secondly for really large objects only a thin-shell near the surface contributes to the fifth force. 

\begin{figure}[htbp]
	\centering
		\includegraphics[width=0.9\columnwidth]{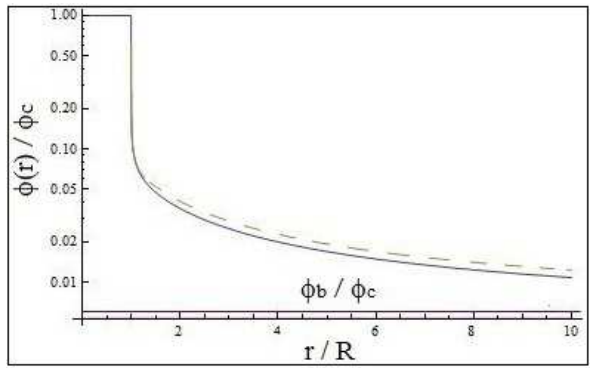}
	\caption{The thin-shell profile for the earth when $(n,k)=(10,1)$ and $m_c R = 10^{6}$ together with the analytical approximation (dashed line). The horizontal line shows $\phi_b$, the minimum in the background. The error between the numerical solution and the analytical approximation is less than $10\%$ in the whole range.}
	\label{fig:thinshell_n10k1a30l73_mcr500k}
\end{figure}

In Fig.~\ref{fig:thinshell_n10k1a30l73_mcr500k} we see the numerical thin-shell profile together with the analytical approximation for $(n,k)=(10,1)$ and $m_c R = 10^6$. To calculate the profile for such highly thin-shelled objects ($m_c R > 100$) it is not possible to start the numerical simulation at $r=0$ since the initial value is too close to $\phi_c$. Upon using the relation (\ref{phiR}) and (\ref{delta_eq}) we are able to start the simulation at $r=R$ allowing us to produce the field profiles shown here.

Note that for large $m_c R$ we have a large gradient at $r=R$ that may cause problems in laboratory experiments  using a very small separation between objects (like Casimir, Eöt-Wash etc.).

\section{The Chameleon force}
The geodesic equation in the Jordan frame reads
\begin{equation}
\ddot{x}^{\mu} + \tilde{\Gamma}^{\mu}_{\alpha\nu}\dot{x}^{\alpha}\dot{x}^{\nu} = 0
\end{equation}
Using $(\ref{csymb})$ this can be rewritten in terms of the Einstein frame connection $\Gamma$ and $\phi$ as
\begin{equation}
\ddot{x}^{\mu} + \Gamma^{\mu}_{\alpha\nu}\dot{x}^{\alpha}\dot{x}^{\nu} = -\beta,_{\phi}\phi^{,\mu}-2\beta,_{\phi}\dot{x}^{\nu}\dot{x}^{\mu}\phi_{,\nu}
\end{equation}
In the non-relativistic limit the last term can be neglected and the chameleon force on a test particle is given by
\begin{equation}
\frac{\vec{F_{\phi}}}{m} = -\beta,_{\phi}\vec{\nabla}\phi
\end{equation}
This is attractive since both $\beta,_{\phi}$ and $\frac{d\phi}{dr}$ are negative  outside a spherical object as shown in section II.
\subsection{Chameleonic Force between two parallel plates}
We consider the force between two identical parallel plates of radius $R_p$ whose surfaces are separated by a distance $d\ll R_p$ and the system is in a laboratory vacuum \cite{Mota:2006fz,Brax:2008hh}. In practice the 'vacuum' will have a non zero pressure corresponding to a very low, but non-zero density $\rho_b$. Because the plates are very close to each other we can treat the plates as infinite flat slabs and take plate 1 to occupy the region $x<-d/2$ and plate 2 to occupy the region $x>d/2$.

Let us  first consider the case when the scalar field is linear with a constant mass $m$ and coupling $\lambda$. Since linearity means the superposition principle holds, we only need to  consider the field emanating from plate 1 in order to calculate the force between the plates. The field-equation for a linear scalar field reads
\begin{align}
	\nabla^2\phi = m^2\phi + \frac{\lambda\rho_c}{M_{\rm pl}}
\end{align}
where $\nabla^2 = \frac{d^2}{dx^2}$ because of the symmetry in the setup. The field equation has the solution
\begin{align}
	\phi(x) = Ce^{mx}-\frac{\lambda\rho_c}{M_{\rm pl} m^2} \text{ for } x<0\\
	\phi(x) = Be^{-mx} \text{ for } x>0
\end{align}
Matching the two solutions at $x=0$ gives us $B=-C=-\frac{\lambda\rho_c}{2M_{\rm pl}m^2}$. The force on the second plate due to the first is then given by
\begin{align}\label{linear_scalar_field}
	\frac{\vert F_{\phi}\vert }{A} &= \frac{\lambda\rho_c}{M_{\rm pl}}\int_{d}^{\infty}\nabla\phi dx = -\frac{\lambda\rho_c}{M_{\rm pl}}\phi(d)\\
	&= 8\pi\lambda^2\frac{G\rho_c^2e^{-md}}{m^2}
\end{align}
where $A$ is the surface area of the plates.
Returning to our model, we  use a subscript $s$ when talking about the quantities defined at the surface of the plates, subscript $b$ in the background and subscript $c$ inside the plates. For example the field-value at the surface of the plates is referred to as $\phi_{s}$, $V(\phi_c)\equiv V_c$ and so on. Also a subscript 0 is used to refer to the quantities where $\dot{\phi}=0$ between the plates. Because of the symmetry this is  at the point $x=0$. Finally we assume that the chameleon mass satisfies $m_c R_p\gg 1$ so that the true non-linear nature of the chameleon comes into play. With the conditions stated at the beginning, we have that $\phi$ obeys
\begin{equation}
\frac{d^2\phi}{dx^2} = V_{,\phi}+\beta,_{\phi}\rho_b
\end{equation}
between the plates, and
\begin{equation}
\frac{d^2\phi}{dx^2} = V_{,\phi}+\beta,_{\phi}\rho_c
\end{equation}
inside either plate. Integrating the equations above yields
\begin{align}\label{casimir_feq}
\begin{array}{cl}
\dot{\phi}^2 = 2(V(\phi)-V_0+\rho_b(\beta(\phi)-\beta_0)) &\text{ for } -d/2<x<d/2\\
\dot{\phi}^2 = 2(V(\phi)-V_c+\rho_c(\beta(\phi)-\beta_c)) &\text{ for } x^2 > d^2/4.
\end{array}
\end{align}
Where we have used that deep inside the plates $\phi(\pm\infty)\approx \phi_c$ and $\frac{d\phi(\pm\infty)}{dx}=0$. Matching at $x=\pm d/2$ we find that the coupling at the surface is given by
\begin{equation}\label{betas}
\beta_s\equiv\beta(\phi_s) = \frac{V_c-V_0 + \rho_c\beta_c-\rho_b\beta_0}{\rho_c-\rho_b}.
\end{equation}
If the second plate were removed $\phi_0 = \phi_b$, the coupling at the surface $\beta_{s0}$ would be given by $(\ref{betas})$, with $\phi_0\to \phi_b$. The perturbation, $\delta\beta_s = \beta_s-\beta_{s0}$, in $\beta(\phi_s)$ due to the presence of the second plate is therefore
\begin{equation}
\delta\beta_s = \frac{V_b-V_0 + \rho_b(\beta_b-\beta_0)}{\rho_c}
\end{equation}
where we have used $\rho_c\gg\rho_b$. Since $m_c R_p\gg 1$ the perturbation deep inside the plates are suppressed exponentially. Using $(1)$ we find that the attractive force on one plate due to the presence of the other one is given by
\begin{equation}
\frac{\vert F_{\phi}\vert }{A} = \rho_c\int_{d/2}^{d/2+D} dx\frac{d\delta\beta(\phi)}{dx} \approx -\rho_c\delta\beta_s
\end{equation}
which, using  $(\ref{betas})$, gives
\begin{equation}
\frac{\vert F_{\phi} \vert}{A} = V_0-V_b+\rho_b(\beta_0-\beta_b) = V_{\text{eff}}(\phi_0)-V_{\text{eff}}(\phi_b)
\end{equation}
We have  to calculate the field value $\phi_0$ halfway between the plates. This is done by integrating ($\ref{casimir_feq}$) over the region $-d/2<x<0$, using that $\frac{d\phi}{dx}<0$ when taking the square root. This gives the equation for $\phi_0$
\begin{equation}
\int_{\phi_0}^{\phi_s}\frac{d\phi}{\sqrt{V(\phi)-V_0+\rho_b(\beta(\phi)-\beta_0)}} = \frac{d}{\sqrt{2}}
\end{equation}
This is a general expression, and can be used for any coupling and potential. Specialising  to our case where $\beta(\phi) = \left(\lambda\frac{H_0}{\phi}\right)^k$ and $V(\phi) = \sigma M^4\left(\frac{\phi}{M}\right)^n$, we change variables to $z = \phi/\phi_0$ and define $z_s = \phi_s/\phi_0$ giving
\begin{equation}\label{phi0int}
\int_{1}^{z_s}\frac{dz}{\sqrt{z^n-1+\frac{n}{k}\left(\frac{\phi_b}{\phi_0}\right)^{n+k}(z^{-k}-1)}} = Md\sqrt{\frac{\sigma}{2}}\left(\frac{\phi_0}{M}\right)^{\frac{n-2}{2}}.
\end{equation}
Here we can have several  cases.
\subsubsection{Case 1: $\phi_0 \approx \phi_c$}
This case corresponds to very small separations $m_b d \ll 1$. We set $\phi_0 = \phi_c(1-\delta)$ and rewrite the right hand side of $(\ref{phi0int})$ as $\frac{m_c d}{\sqrt{2n(n+k)}}$. The integral can now be evaluated
\begin{equation}
\int_{1}^{1+\delta}\frac{dz}{\sqrt{n(z-1)}} = \frac{2\sqrt{\delta}}{\sqrt{n}}
\end{equation}
resulting in
\begin{equation}
\delta = \frac{(m_c d)^2}{8(n+k)}.
\end{equation}
This case only applies when the separation $d$ is much smaller than the thickness of the plates $t$, since we have assumed $m_c t\gg 1$. The chameleon force becomes
\begin{equation}
\frac{\vert F_{\phi}\vert }{A} = V_c\left[1 - \frac{n(m_c d)^2}{8(n+k)}\right]
\end{equation}
\subsubsection{Case 2: $\phi_0 \approx \phi_b$}
This corresponds to the case when the field drops all the way down to the minimum in between the bodies. Since this case corresponds to $m_b R > m_b d > 1$ the force is  exponentially suppressed. We put $\phi_0 = \phi_b(1+\delta)$ where we assume $\delta \ll 1$. This allows us to approximate $\left(\frac{\phi_b}{\phi_0}\right)^{n+k}\approx 1 -(n+k)\delta$ and since $\phi_c \gg \phi_b$ we can take $z_s\to \infty$ and the integral (\ref{phi0int}) can be written
\begin{equation}
\int_{1}^{\infty}\frac{dz}{\sqrt{z^n-1+\frac{n}{k}\left(1-(n+k)\delta\right)(\frac{1}{z^k}-1)}} = \frac{m_b d}{\sqrt{2n(n+k)}}.
\end{equation}
In the limit $\delta\to 0$ the left hand side diverges. Upon using a power series expansion of the integrand near $z=1$
\begin{eqnarray}
&z^n- 1 + \frac{n}{k}\left(\frac{\phi_b}{\phi_0}\right)^{n+k}(z^{-k}-1) \approx n(n+k)  \times \\   \nonumber& \left[\delta(z-1) + \frac{1}{2}\left(1-\delta(k+1)\right)(z-1)^2\right.\\ &\left.+\frac{(n-k-3)-(k+1)(k+2)\delta}{6}(z-1)^3 +...\right]
\end{eqnarray}
we see that the second term is the divergent part when $\delta=0$. This term dominates in the region $1+2\delta<z<\left.|\frac{n-k}{n-k-3}\right|$ and  for $0<\delta\ll 1$ provides the dominating contribution to the integral. We can therefore approximate the integral by
\begin{equation}
\int_{1+2\delta}^{\left.|\frac{n-k}{n-k-3}\right|}\frac{dz}{\sqrt{\frac{n(n+k)}{2}}(z-1)} \approx \frac{\sqrt{2}\ln(2\delta)}{\sqrt{n(n+k)}}
\end{equation}
This gives
\begin{equation}
\delta \approx \frac{1}{2}e^{-\frac{m_b d}{2}}
\end{equation}
and shows that the chameleon force
\begin{equation}
\frac{\vert F_{\phi}\vert }{A} \approx V_{\rm eff,\phi\phi}(\phi_b)\frac{(\phi_0-\phi_b)^2}{2} \approx \frac{m_b^2\phi_b^2}{8}e^{-m_b d}
\end{equation}
is indeed exponentially suppressed by the factor $m_b d\gg 1$.
\subsubsection{Case 3: $\phi_c\gg\phi_0\gg\phi_b$}
\begin{figure}
\begin{center}
\includegraphics[width=0.9\columnwidth]{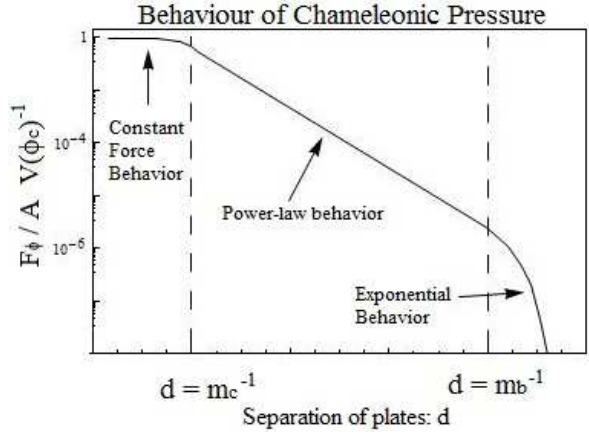}
\caption{General behaviour of the chameleon pressure $\frac{F_{\phi}}{A}$ as a function of the plate separation $d$.}%
\label{fig:genchamforce}
\end{center}
\end{figure}

In this last case we can neglect the third term in the square root of $(\ref{phi0int})$ and also take $z_s\to\infty$. This enables us  to evaluate the integral analytically
\begin{equation}
\int_{1}^{\infty}\frac{dz}{\sqrt{z^n-1}} = \frac{\Gamma\left(\frac{1}{2}\right)\Gamma\left(\frac{1}{2}-\frac{1}{n}\right)}{|\Gamma\left(-\frac{1}{n}\right)|}.
\end{equation}
The $\Gamma$-function satisfies $\Gamma(\epsilon) \approx \frac{1}{\epsilon}-\gamma_{E}$ for $\epsilon\ll 1$ with $\gamma_{E}\approx 0.577$ being the Euler-Gamma constant. This gives
\begin{equation}
S_n \equiv \frac{\Gamma\left(\frac{1}{2}\right)\Gamma\left(\frac{1}{2}-\frac{1}{n}\right)}{|\Gamma\left(-\frac{1}{n}\right)|} \approx \frac{\pi}{n} \text{ for large  } n.
\end{equation}
We can now find an explicit expression for $\phi_0$
\begin{equation}
\phi_0 = M\left(\sqrt{\frac{2}{\sigma}}\frac{S_n}{Md}\right)^{\frac{2}{n-2}}
\end{equation}
and the chameleon force
\begin{align}
\frac{\vert F_{\phi}\vert }{A} \approx \sigma M^4\left(\sqrt{\frac{2}{\sigma}}\frac{S_n}{Md}\right)^{\frac{2n}{n-2}}.
\end{align}
We see that the force follows a power-law where the drop-off is faster than $1/d^2$, but slower than $1/d^4$ when $n>4$. The Casimir-force falls off as $1/d^4$ making Casimir experiments (with large plate-separations) a powerful way of constraining the chameleon. See Fig.~\ref{fig:genchamforce} for a plot of the chameleonic force (or more accurately, the pressure $F_{\phi}/A$) as a function of the distance between the plates.

\subsection{Chameleon Force between two spherical thin-shelled bodies}
We consider the force between two bodies, with thin-shells, that are separated by a distance $r\gg R_1,R_2$. Given than $r\gg R_1, R_2$ we can consider the monopole moment of the field emanating from the two bodies only.

We denote by $\phi_1$ ($\phi_2$)  the field outside body one (two) when body two (one) is absent. To a good accuracy we have $\phi_1 \approx \phi_{c1}$ the minimum inside body 1. In between the bodies, we can superimpose the far-away  fields from the two bodies. As the distance is large,  the perturbation $\delta\phi_1$ in the field inside the body two due to the presence of body one  satisfies $\delta\phi_1 \ll \phi_2$. The combined field close to the surface of body two is approximately given by $\phi_2+\delta\phi_1$.

Using the geodesic equation, $dF_{\phi} = -\beta,_{\phi}\nabla\phi dm$, we have that the total force on body two due to body one is
\begin{align}
	\vert F_{\phi}\vert  \approx \beta,_{\phi_{c2}}\int_{\text{body two}}\nabla\delta\phi_1dm
\end{align}
Next we have that the perturbation $\delta\phi_1$ is given by the far field of body one evaluated at body two
\begin{align}
	\delta\phi_1 = \frac{\beta_{\text{eff1}}}{4\pi M_{\rm pl}}\frac{M_1e^{-m_br}}{r}
\end{align}
Because of the big mass of the chameleon inside body two, the perturbation created by body one is attenuated, and, as in the SCM, only a thin-shell close to the surface contributes to the force. We model this by setting
\begin{align}
	\vert F_{\phi}\vert  = 2\beta_{\text{eff}1}\beta,_{\phi_{c2}}M_{\rm pl}\left(\frac{\Delta R}{R}\right)_2\frac{GM_1M_2(1+m_br)e^{-m_br}}{r^2}
\end{align}
where $\left(\frac{\Delta R}{R}\right)_2$ models the effect of this thin-shell.

Likewise the force on body one due to body two is given by the same expression with $1\to 2$. Up to a $\mathcal{O}(1)$ factor we have
\begin{align}
	\frac{\Delta R}{R} = \frac{\phi_c}{\rho_c|\beta,_{\phi_c}|R^2(m_cR)^{\frac{2}{n-2}}}\sim \frac{1}{(m_cR)^{\frac{2n-2}{n-2}}}
\end{align}
which equals $\frac{\beta_{\text{eff}}}{|\beta,_{\phi_c}|M_{\rm pl}}$ up to a $\mathcal{O}(1)$ factor. The force between two thin-shelled objects is then given by
\begin{align}\label{forcelaw}
\vert F_{\phi}\vert = 2\beta_{\text{eff}1}\beta_{\text{eff}2}\frac{GM_1 M_2(1+m_br)e^{-m_br}}{r^2}
\end{align}
where we have chosen an appropriate $\mathcal{O}(1)$ factor. In the thick-shell case ($m_c R \ll 1$) the whole body contributes to the force\footnote{The field-equation is quasi-linear and the superposition principle holds.} giving
\begin{align}
	\vert F_{\phi}\vert = 2(\beta,_{\phi_i}^{(1)}M_{\rm pl})(\beta,_{\phi_i}^{(2)}M_{\rm pl})\frac{GM_1M_2(1+m_br)e^{-m_br}}{r}
\end{align}
\section{Bounds on the parameters}
We will constrain the parameters $\lambda$ and $M$ (or $\sigma$) by looking at the consequences our model has on local gravity experiments. The experiments considered here  restricts the value of  the chameleon coupling in different regions. The Eöt-Wash experiment (and other fifth-force searches) are usually the best way to obtain good bounds when $|\beta,_{\phi_c}|M_{\rm pl}\sim 1$. Casimir type experiments are often the best way to bound the highly coupled, $|\beta,_{\phi_c}|M_{\rm pl} \gg 1$, region. Finally the PPN and BBN bounds constrain the extremely high coupled region which are invisible to the Casimir type experiments due to the extremely short range of the chameleon.
\subsection{PPN bounds}
\begin{figure}%
\centering
\includegraphics[width=0.8\columnwidth]{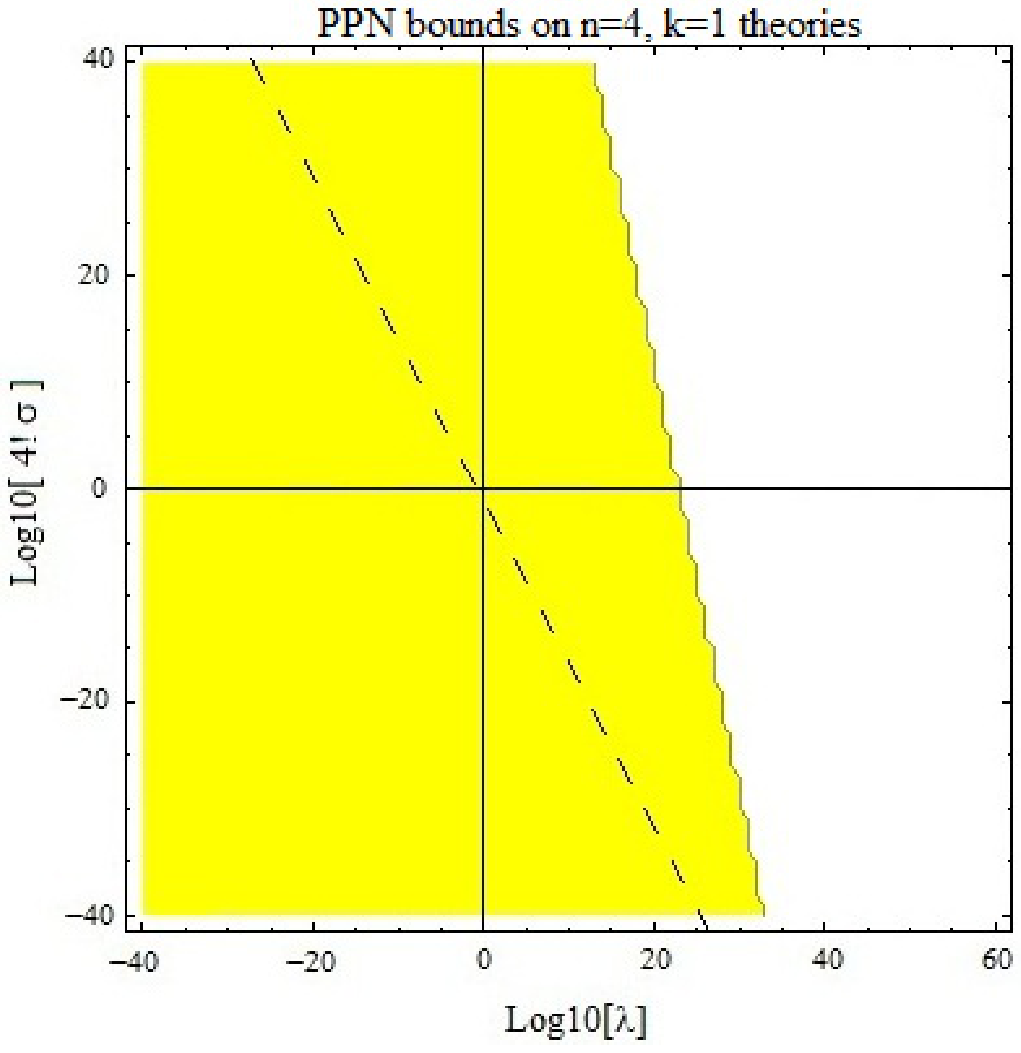}\\
\includegraphics[width=0.8\columnwidth]{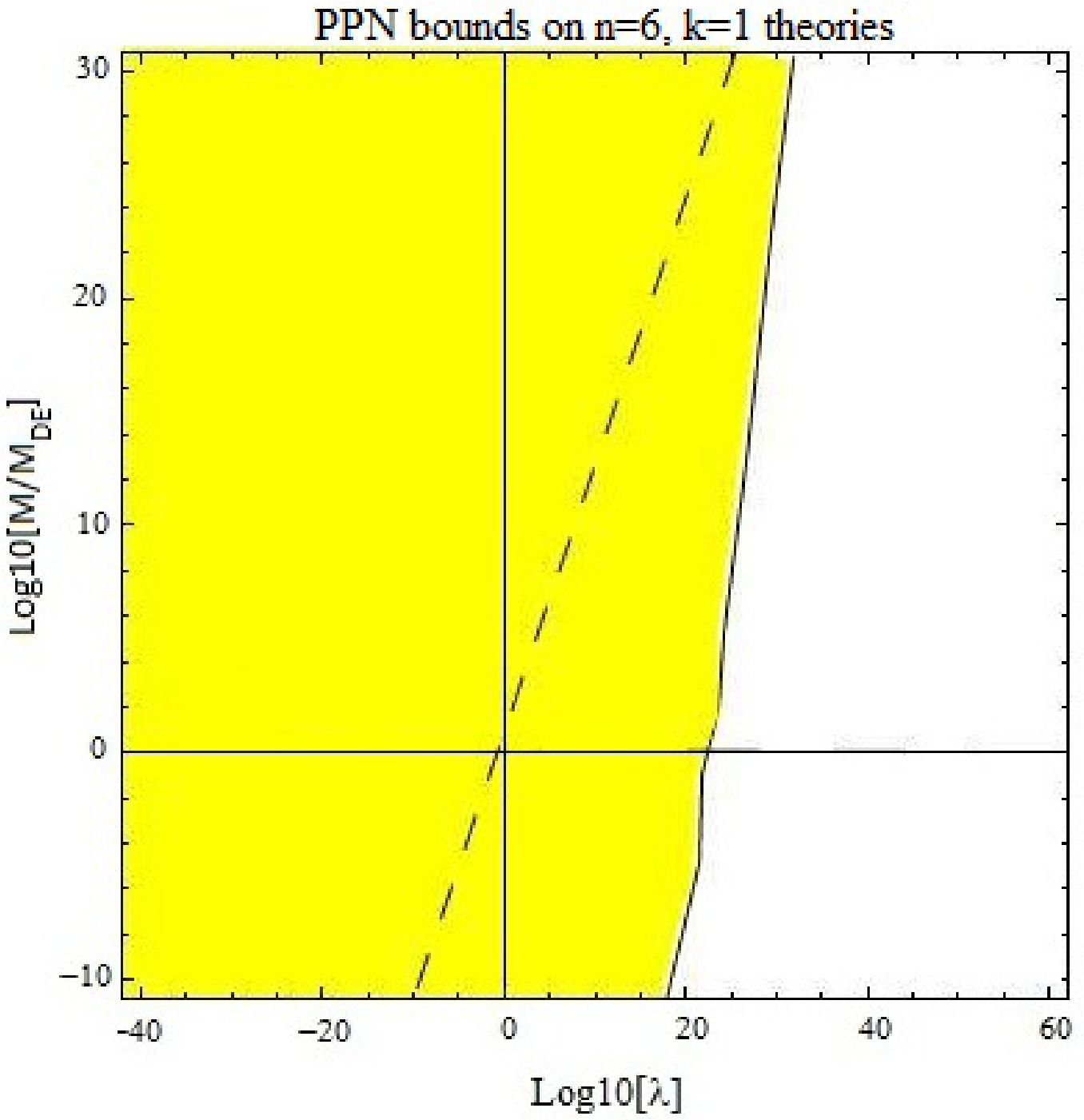}\\
\includegraphics[width=0.8\columnwidth]{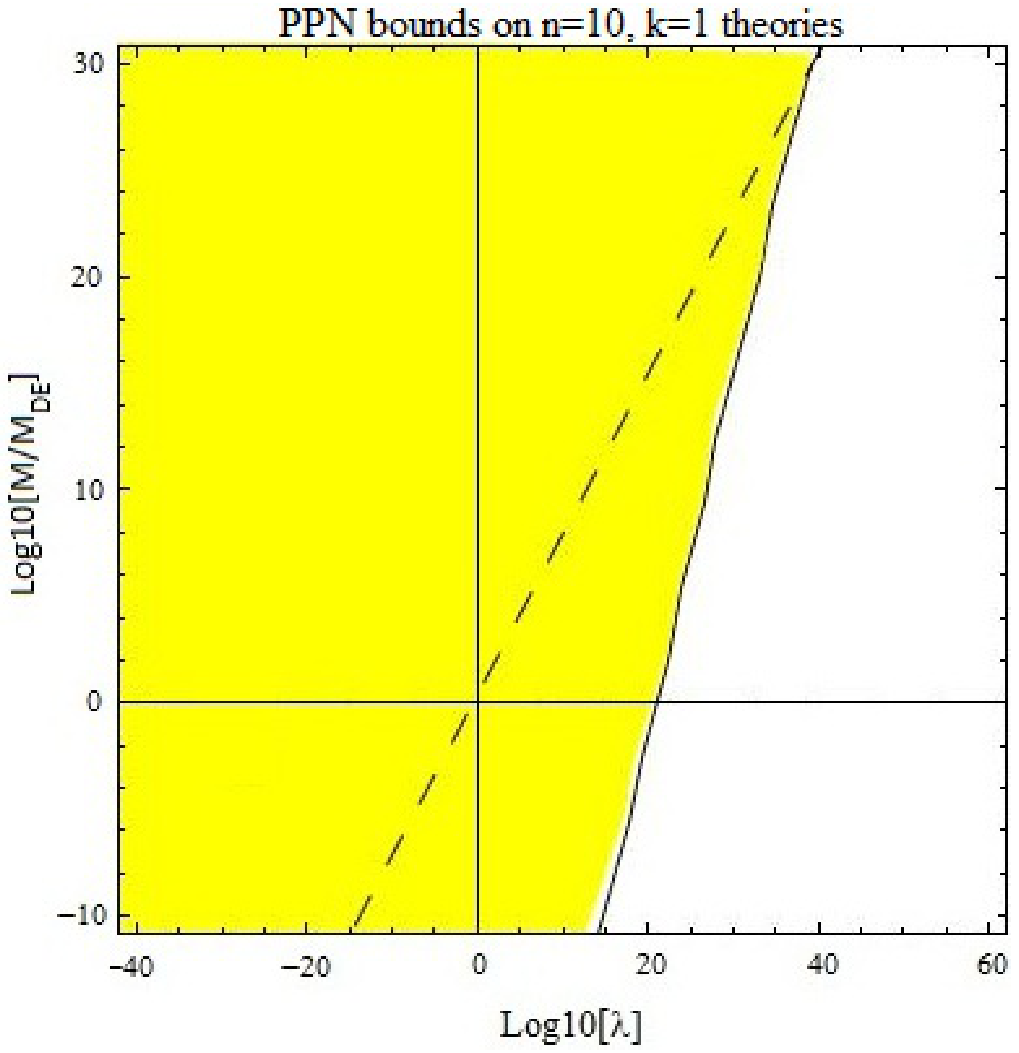}%
\caption{PPN constraints on chameleon theories coming from experimental bounds on the Eddington-parameter in light-deflection experiments. The shaded area shows the regions of parameter space that are allowed by the current data. The solid horizontal black lines indicate the cases where $M$ and $\sigma$ take 'natural values'. The solid vertical lines show when $M_{\beta} = H_0$. The dashed black line indicates when $|\beta,_{\phi_c}|M_{\rm pl}=1$ for $\rho_c = \mathcal{O}(1g/cm^3)$. The amount of allowed parameter space increases with $n$.}%
\label{PPNbounds}%
\end{figure}

For experiments using the deflection of light by large bodies, the only Post-Newtonian Parameter (PPN) at play is the Eddington-parameter $\gamma$. The Eddington-parameter is defined in the Jordan-frame by $\tilde{g}_{ij} = (1-2\gamma\tilde{\Psi})\delta_{ij}$ when $\tilde{g}_{00} = -1-2\tilde{\Psi}$ \cite{Hinterbichler:2010es}. Transforming to the Einstein-frame we get the following estimate for $\gamma$
\begin{equation}
\gamma = \frac{\Psi_E-\beta(\phi)}{\Psi_E+\beta(\phi)} \approx 1 - \frac{2\beta(\phi)}{\Psi_E}
\end{equation}
The back reaction on the gravitational potential from the chameleon is in most interesting cases negligible, and since $\beta(\phi)\ll 1$ the Jordan-frame and Einstein-frame potential are the same. The best bounds on this parameter comes from the Cassini-experiment \cite{Bertotti:2003rm} and reads $|\gamma-1| < 2.3\cdot 10^{-5}$. The gravitational potential for the sun is $\Psi_{\rm sun} = 10^{-6}$ and the field near the surface of the sun satisfies $\phi\approx \phi_c$ giving us the bounds shown in Fig.~\ref{PPNbounds}. This experiment only restricts the parameters in which $|\beta,_{\phi_c}|M_{\rm pl}\gg 1$.
\subsection{BBN bounds}
Since our chameleon couples to matter via the conformal transformation (\ref{conformal_transform}), the masses of the standard model particles  have a $\phi$-dependence of the form $m=m_{0}\exp{\beta(\phi)}$. Bounds on particle masses restrict a variation of this type to be below the 10$\%$ level since  Big-Bang Nucleosynthesis (BBN) \cite{Brax:2004qh}. Since in our model $\dot{\phi}<0$, $\beta(\phi)$ is an increasing function of time so we must require
\begin{equation}
\beta(\phi_{\rm today})\lesssim 0.1 \text{ and } \beta(\phi_{\rm BBN})\lesssim 0.1
\end{equation}
The last condition is satisfied as long as the chameleon has settled at the minimum before the time of BBN. The condition today translates into the bound
\begin{align}
\lambda \lesssim 10^{30}\left(\frac{M}{M_{DE}}\right)^{\frac{n-4}{n}} \text{ for } n\not=4 \\
\lambda \lesssim 10^{30}\sigma^{-\frac{1}{4}} \text{ for } n=4
\end{align}
The $k$ dependence is weak, and we have that this bound is satisfied as long as the PPN bound above is satisfied.
\subsection{Eöt-Wash bounds}
The University of Washington Eöt-Wash experiment \cite{Adelberger:2002ic} is designed to search for deviations from the $1/r^2$
drop-off of Newton's law. The experiment uses a rotating torsion balance to measure the
torque on a pendulum. The torque on the pendulum is induced by an attractor which rotates with a frequency $\omega$.
The attractor has 42 equally spaced holes, or missing masses, bored into it. As a result, any torque on the pendulum,
which is produced by the attractor, will have a characteristic frequency which is some integer multiple of 21$\omega$. This
characteristic frequency allows any torque due to background forces to be identified.  The attractor is manufactured so that, if
gravity drops off as $1/r^2$, the torque on the pendulum vanishes.
The experiment has been running  with different separations between the pendulum and the attractor. The experiment has been running for typically separations $d=55\mu m$. Both the attractor and the pendulum are made out of molybdenum with a density of about $\rho_c = 10 g/cm^3$ and are $t=0.997mm$ thick.
Electrostatic forces are shielded by placing a $d_{\rm shield}=10 \mu m$ thick, uniform BeCu sheet between the attractor and the pendulum.
The density of this sheet is $\rho_{\rm shield}= 8.4 g/cm^3$. As discussed in \cite{Mota:2006fz} the role played by this sheet is crucial when testing for chameleon fields in the strong coupling regime. If the coupling is strong enough, the sheet will itself develop a thin-shell. When this occurs the effect of the sheet is not
only to shield electrostatic forces, but also to block any chameleon force originating from the attractor. Following the analogy of our model with the SCM this effect is given by an extra suppression of $e^{-m_{\rm shield}d_{\rm shield}}$. And, in effect, this will make a larger part of the parameter space allowed in the strongly coupled case. It will not affect the experiment when $|\beta,_{\phi_c}|M_{\rm pl}\sim 1$. The force per unit area between the attractor and the pendulum plates due to a scalar field with matter coupling $\lambda$ and constant mass $m$, where $1/m \ll 0.997mm$ is given by (\ref{linear_scalar_field})
\begin{align}
	\frac{\vert F_{\phi}\vert }{A} = \alpha \frac{G\rho_c^2e^{-md}}{2m^2}
\end{align}
where $\alpha = 8\pi\lambda^2$ and $d$ is the separation of the two plates. The strongest bound on $\alpha$ coming from the Eöt-Wash
experiment is $\alpha < 2.5 \cdot 10^{-3}$ for $1/m = 0.4-0.8mm$ which constrains $\lambda < 10^{-2}$.
\\\\
When the pendulum and the attractor have thin-shells the force is given by the expressions derived in  section III. The vacuum used in these experiments
has a pressure of $p=10^{-6}$ Torr which means that the chameleon mass in the background, $m_b$, is non-zero and for the largest couplings we will have a  $e^{-m_b d}$ suppression. Hence the experiment cannot detect a very strongly coupled chameleon. The BeCu sheet produces a force on the pendulum. As the sheet is uniform, this  resulting force leads to no detectable torque. If neither the pendulum nor the attractor have thin-shells then we must have $m_bd \ll 1$ and the chameleon force is simply $2\beta,_{\phi_i}^2 M_{\rm pl}^2$ times the gravitational one. Since this force drops off as $1/r^2$, it will be undetectable in this experiment. In this case, however, the model-parameters are constrained by other experiments such as those that look for Yukawa forces with larger ranges as discussed below.

\begin{figure}%
\centering
\includegraphics[width=0.8\columnwidth]{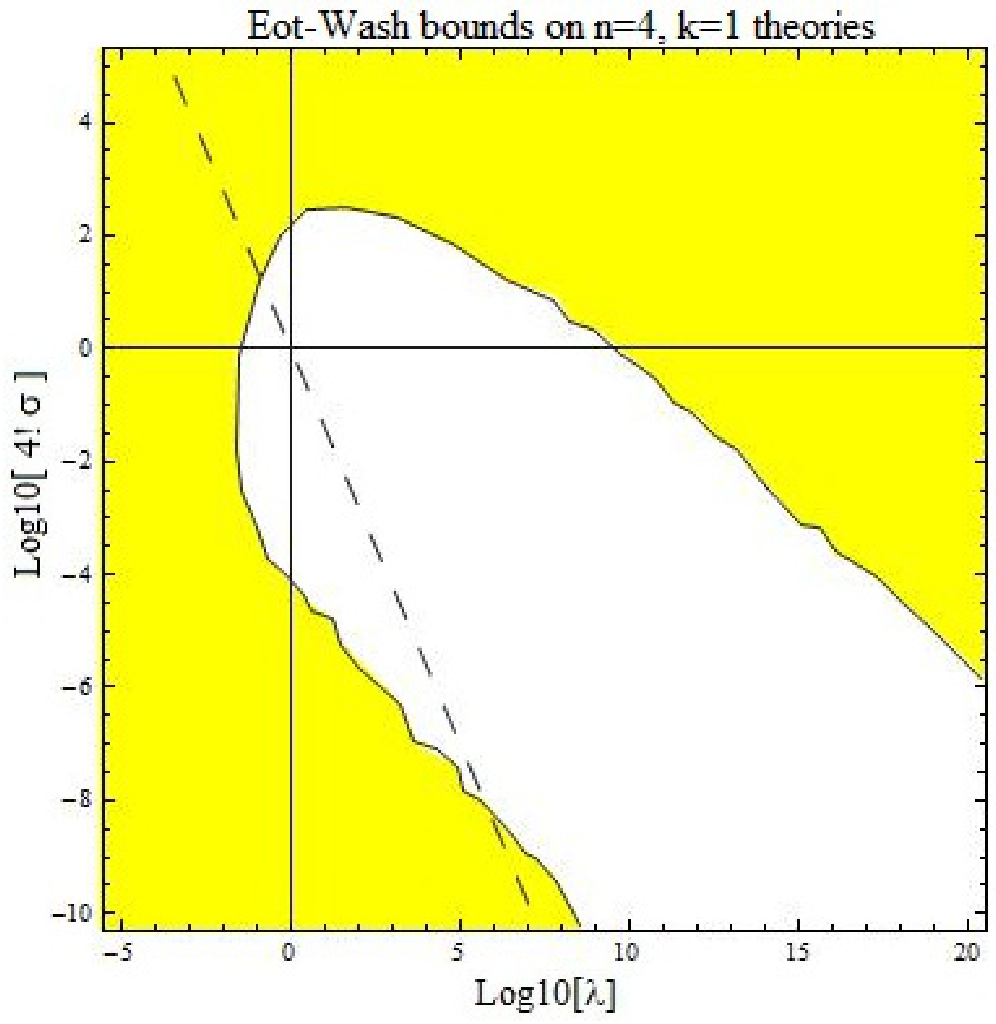}\\
\includegraphics[width=0.8\columnwidth]{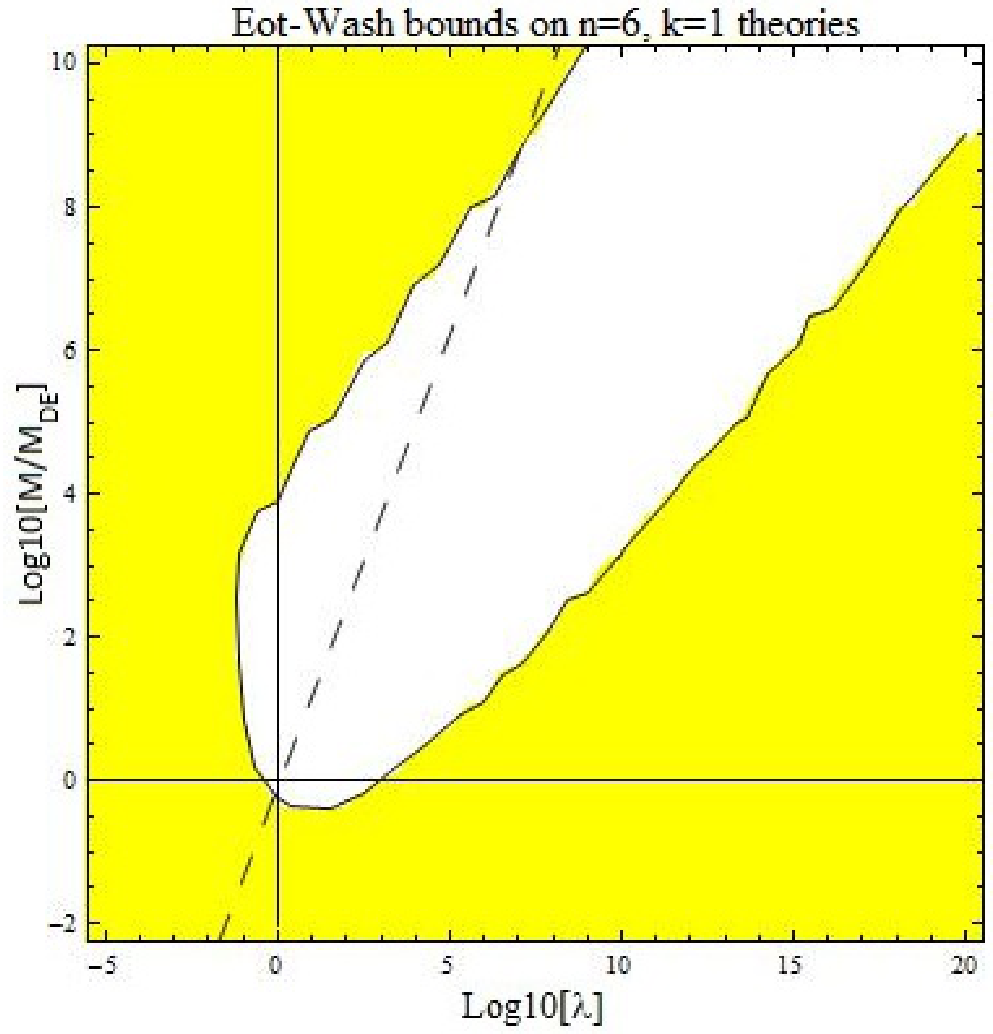}\\
\includegraphics[width=0.8\columnwidth]{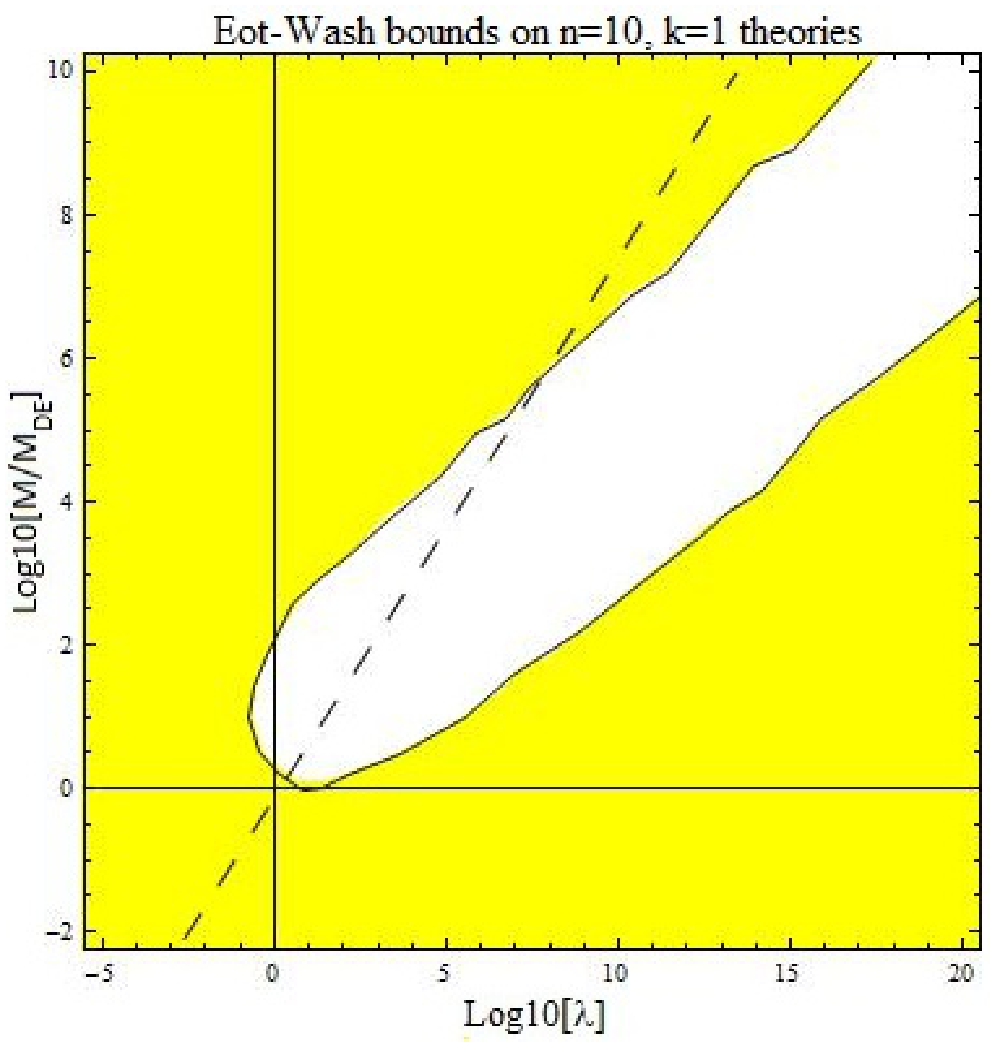}
\caption{Constraints on chameleon theories coming from Eöt-Wash bounds on deviations from Newton's
law. The shaded area shows the regions of parameter space that are allowed by the current data. The solid horizontal black lines indicate the cases where $M$ and $\sigma$ take 'natural values'. The solid vertical lines show when $M_{\beta} = H_0$. The dashed black line indicates when $|\beta,_{\phi_c}|M_{\rm pl}=1$ for $\rho_c = \mathcal{O}(1g/cm^3)$. The amount of allowed parameter space increases with $n$.}%
\label{eotwashbounds}%
\end{figure}

Even though we have formulae for the force, we have used numerics to calculate the bounds. This  gives  more accuracy in the regions where our approximate formulae do not apply. The torque generated by the rotation of the plates can be shown to be given by \cite{Brax:2008hh}
\begin{equation}
\tau_{\phi} \approx e^{-m_{\text{shield}}d_{\text{shield}}}a_T\int_d^{\infty}\frac{F_{\phi}(x)}{A}dx
\end{equation}
where $a_T = \frac{dA}{d\theta}$ is a constant that depends on the experimental setup and the exponential models the effect of the electrostatic shield. For the 2006 E$\ddot{\text{o}}$t-Wash experiment $a_T = 3\cdot 10^{-3}m^2$. The bounds derived from the experiment can also be expressed in terms of this torque as $\tau_{\phi}(d=55\mu m) < 0.87\cdot 10^{-17}$Nm, which we have used to compute the bounds numerically. We have also compared the numerical results and  the analytical expression in the regions where they both apply. Our results are shown in Fig.~\ref{eotwashbounds}. In these plots the shaded region is allowed by the current bounds.

When $n = 4$, we can see that a natural value of $\sigma$ is ruled out for $\lambda = 1$. As $n$ becomes larger than 8, the case $\lambda \sim 1$ becomes allowed for $M = M_{DE}$. There exists, for each $n$, a large region of the parameter space which is allowed by the experiment and in which $|\beta,_{\phi_c}M_p|\gg 1$.

The area of allowed parameter space grows with increasing $n$. Indeed when the potential is steeper, the mass of the chameleon increases, and the thin-shell effect is present for a larger part of the parameter space.

The setup and the behaviour of a chameleon in the experiment is more thoroughly explained in \cite{Brax:2008hh}.
\subsection{Fifth-force searches}
\begin{figure}%
\centering
\includegraphics[width=0.8\columnwidth]{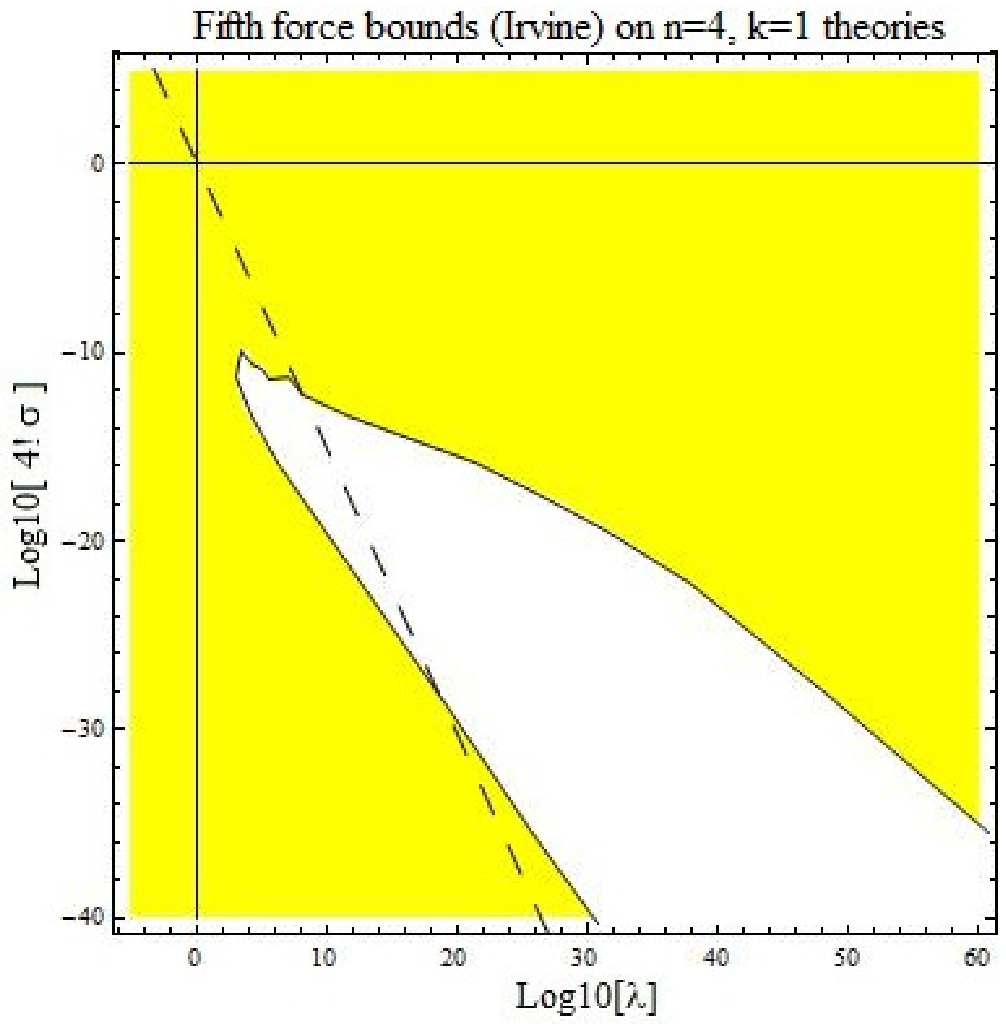}\\
\includegraphics[width=0.8\columnwidth]{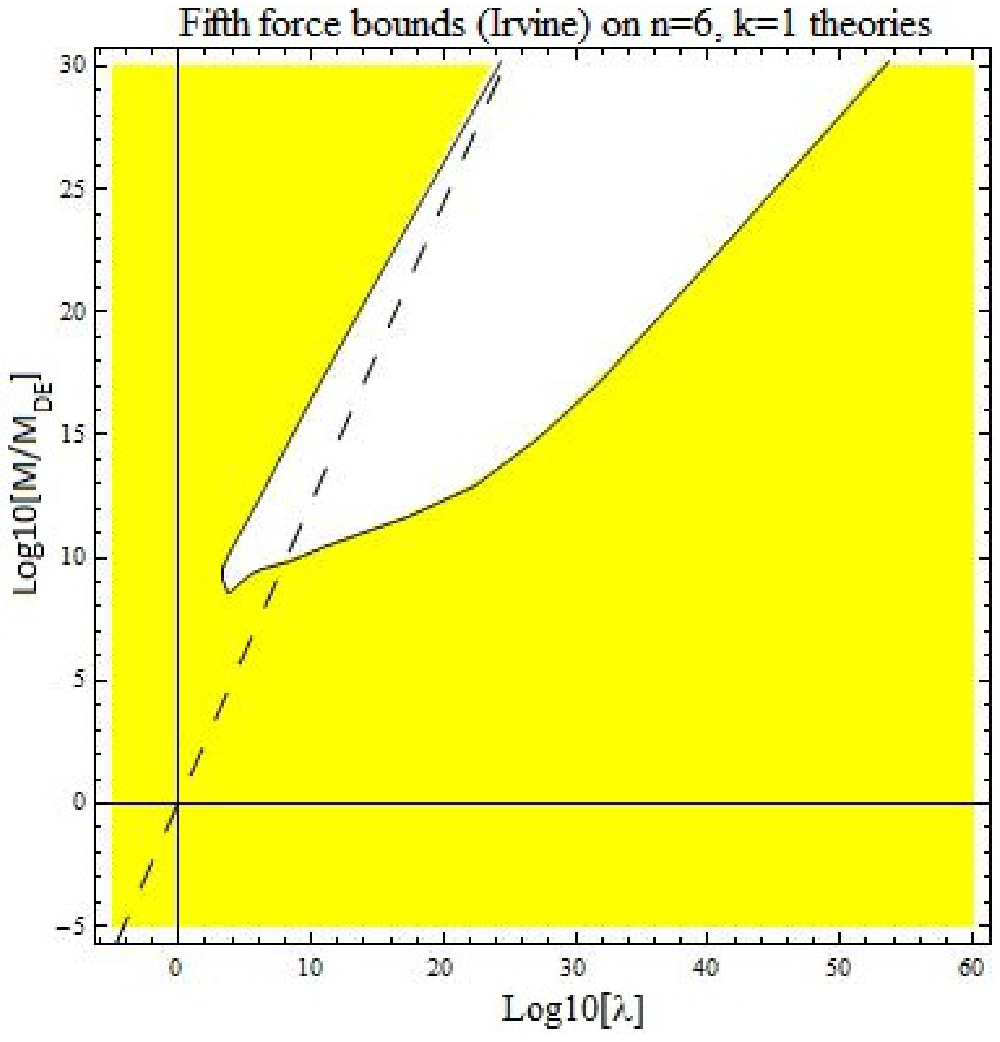}\\
\includegraphics[width=0.8\columnwidth]{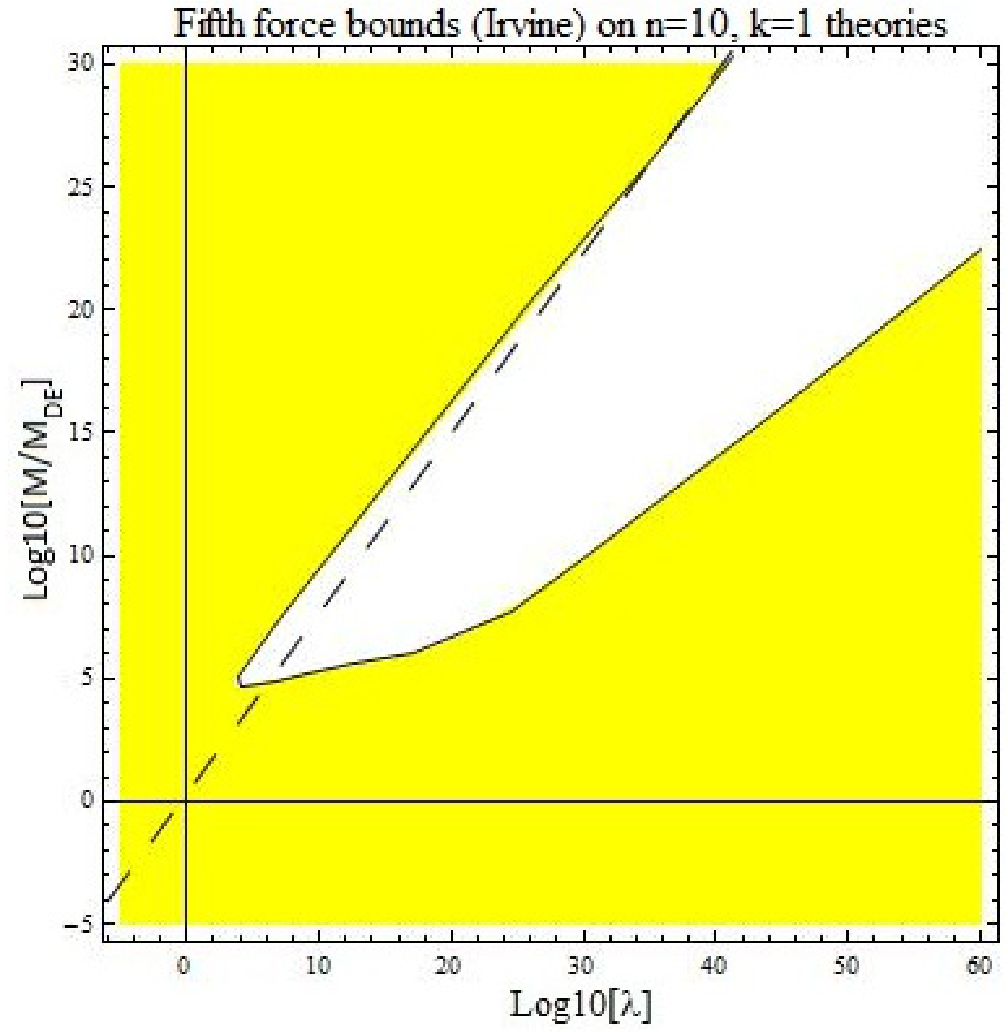}
\caption{Constraints on chameleon theories coming from experimental fifth-force searches (the Irvine-experiment). The shaded area shows the regions of parameter space that are allowed by the current data. The solid horizontal black lines indicate the cases where $M$ and $\sigma$ take 'natural values'. The solid vertical lines show when $M_{\beta} = H_0$. The dashed black line indicates when $|\beta,_{\phi_c}|M_{\rm pl}=1$ for $\rho_c = \mathcal{O}(1g/cm^3)$.}%
\label{Fifthforce}%
\end{figure}

In the Irvine-experiment \cite{Hoskins:1985tn} the inverse-square distance dependence of the Newtonian gravitational force law was tested. One experiment used a torsion balance consisting of a 60-cm-long copper bar suspended at its midpoint by a tungsten wire, to compare the torque produced by copper masses 105 cm from the balance axis with the torque produced by a copper mass 5 cm from the side of the balance bar, near its end. The produced torques due to the masses at 105 cm and 5 cm have been  measured. Letting $R_{\rm Measured}$ be the measured ratio of the two torques and $R_{\rm Newton}$ the Newtonian prediction it was found that
\begin{align}
	\left|\frac{R_{\rm Measured}}{R_{\rm Newton}}-1\right| = (1.2\pm 7) \cdot 10^{-4}
\end{align}
If the walls of the vacuum chamber do not have thin-shells, the field inside the chamber, as discussed below (\ref{mieq}), settles at a value where $m_{\text{chamber}}\sim R_{\text{chamber}}^{-1}$ where $R_{\text{chamber}}$ is the size of the chamber. The experiment here bounds
\begin{align}
2\beta,_{\phi_i^{(1)}}\beta,_{\phi_i^{(2)}}M_{\rm pl}^2 \lesssim 10^{-3}
\end{align}
with $\phi_i$ determined by (\ref{phiieq}). The vacuum chamber used was held at a pressure $p=3\cdot 10^{-8}$torr which corresponds to a background density $4.6\cdot 10^{-14} g/cm^3$ (at $T=300K$). When the walls of the chamber (and therefore the test-masses) have thin-shells the chameleon sits at the minimum of its effective potential inside the chamber. The chameleon mass $m_{\rm chamber}$ is typically much larger than the inverse size of the chamber and the bound becomes
\begin{align}
2\beta_{\text{eff1}}\beta_{\text{eff2}} \lesssim 10^{-3}
\end{align}
where $\beta_{\text{eff}}$ is the thin-shell effective coupling given by (\ref{eq:betaeff}). For the highly coupled cases $m_{\text{chamber}} R_{{chamber}} \gg 1$ there is an extra $e^{-m_{\text{chamber}}d}$ suppression of the torque where $d$ is the separation of the test-masses. This experiment provides the best bounds for the chameleon in the linear regime since the more accurate Eöt-Wash experiment is, by design, unable to detect the linear chameleon ($F_{\phi}\propto 1/r^2$). See Fig.~\ref{Fifthforce} for the resulting bounds.
\subsection{Casimir bounds}
Casimir force experiments provide an excellent way of  bounding  chameleon field parameters when  the scalar field is strongly coupled to matter. Casimir force experiments measure the force per unit area between two test masses separated by a distance $d$. It is generally the case that $d$ is small compared to the curvature of the surface of the two bodies and so the test masses can be modeled, to a good approximation, as flat plates and the results derived in section III apply. The Casimir force between two parallel plates is:
\begin{align}
	\frac{\vert F_{\rm Casimir}\vert }{A} = \frac{\pi^2}{240d^4}
\end{align}
Even though the most accurate measurements of the Casimir force have been made using one sphere and one slab as the test bodies, this setup has a more complicated geometry and will not be discussed in this paper. We will focus on the experiments which use two flat slabs as test bodies.

In all cases, apart from  $n = 4$ and $m_cd\gg 1$, the chameleon force per area grows more slowly than $d^4$ as
$d \to 0$. When $n = 4$ and $m_cd\gg 1$, $m_bd\ll 1$ we have $F_{\phi}/A\propto d^{-4}$. It follows that the larger the separation, $d$,  the better Casimir force searches constrain chameleon theories. Additionally, these tests provide the best bounds when
the test masses do have thin-shells as this results in a strongly $d$ dependent chameleon force.

Note that if the background chameleon mass is large enough $m_bd \gg 1$ then $F_{\phi}/A$ is suppressed by a factor $e^{-m_bd}$. This shows that the experiments cannot detect the strongest coupled chameleons. For these extreme cases the post-newtonian corrections (and BBN bounds)  constrain these theories. See \cite{Brax:2007vm} for a detailed analysis of the Casimir force in the SCM.

\begin{figure}%
\centering
\includegraphics[width=0.8\columnwidth]{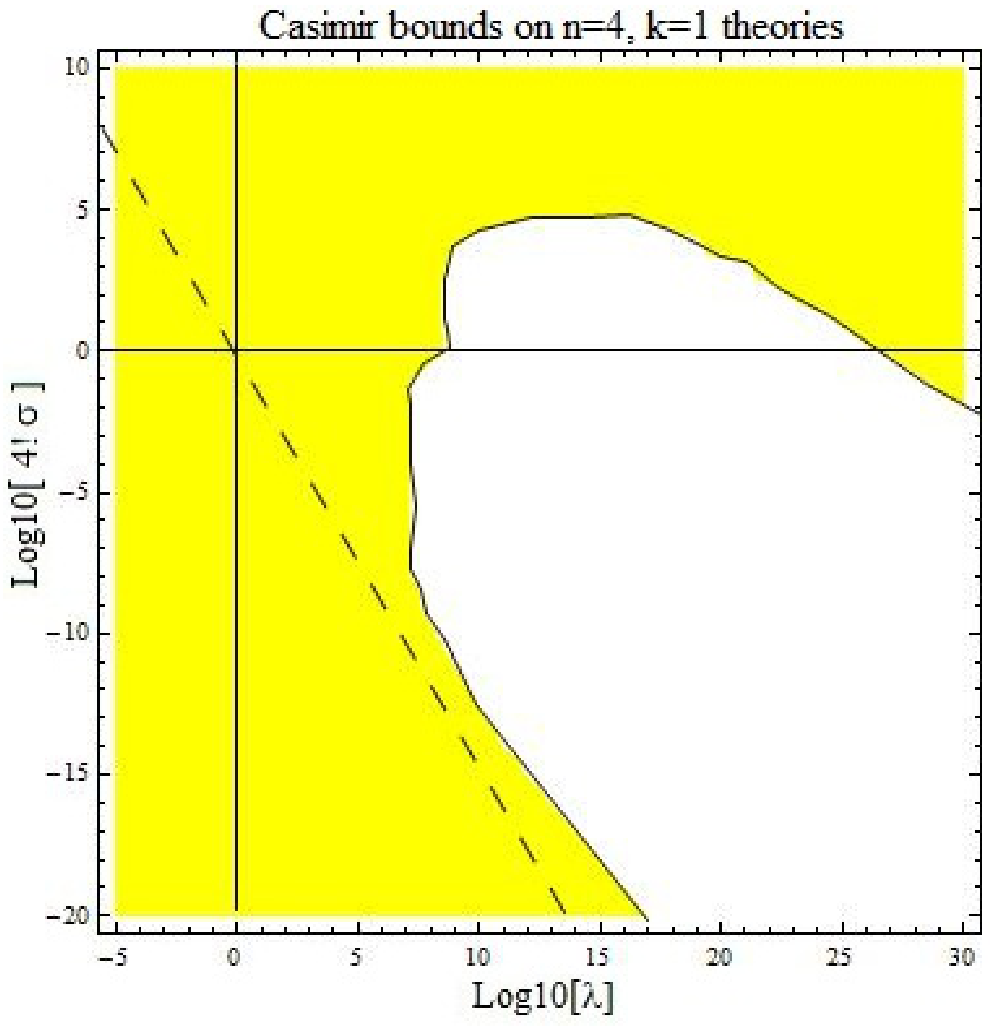}\\
\includegraphics[width=0.8\columnwidth]{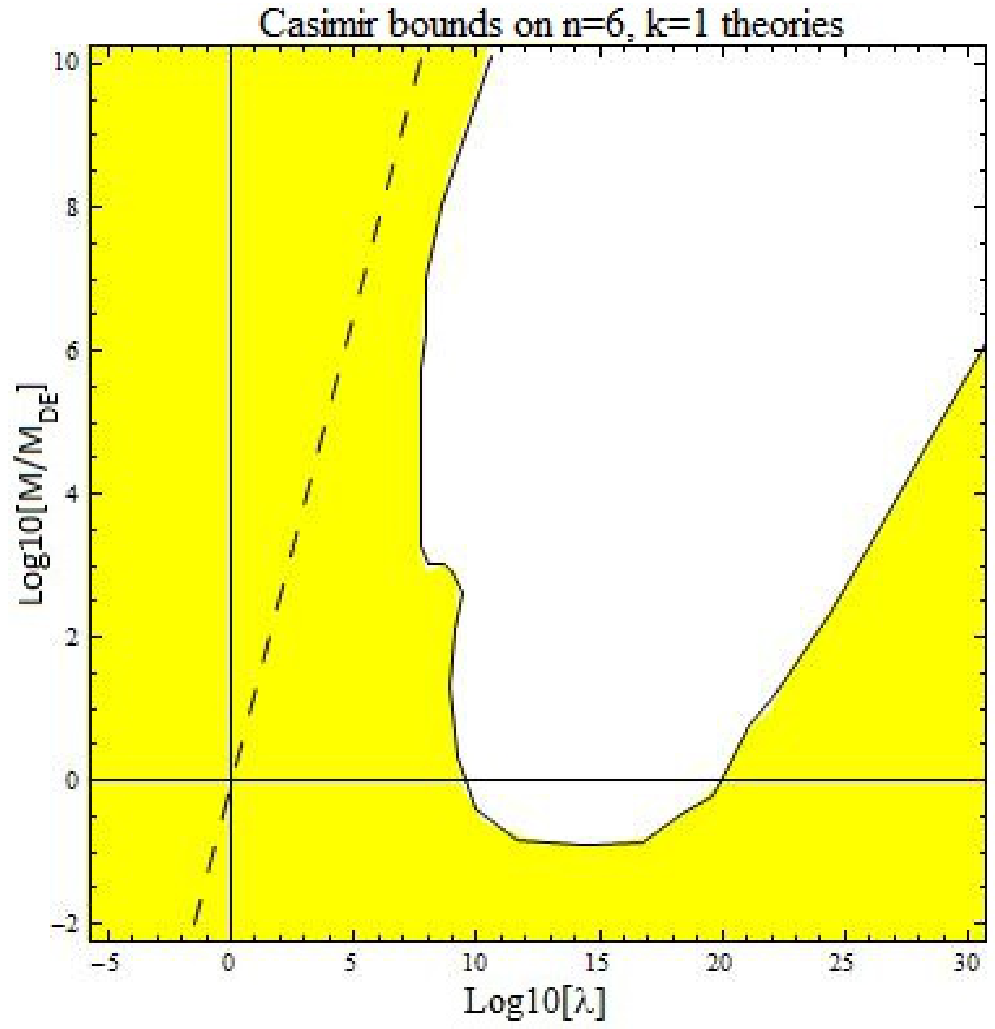}\\
\includegraphics[width=0.8\columnwidth]{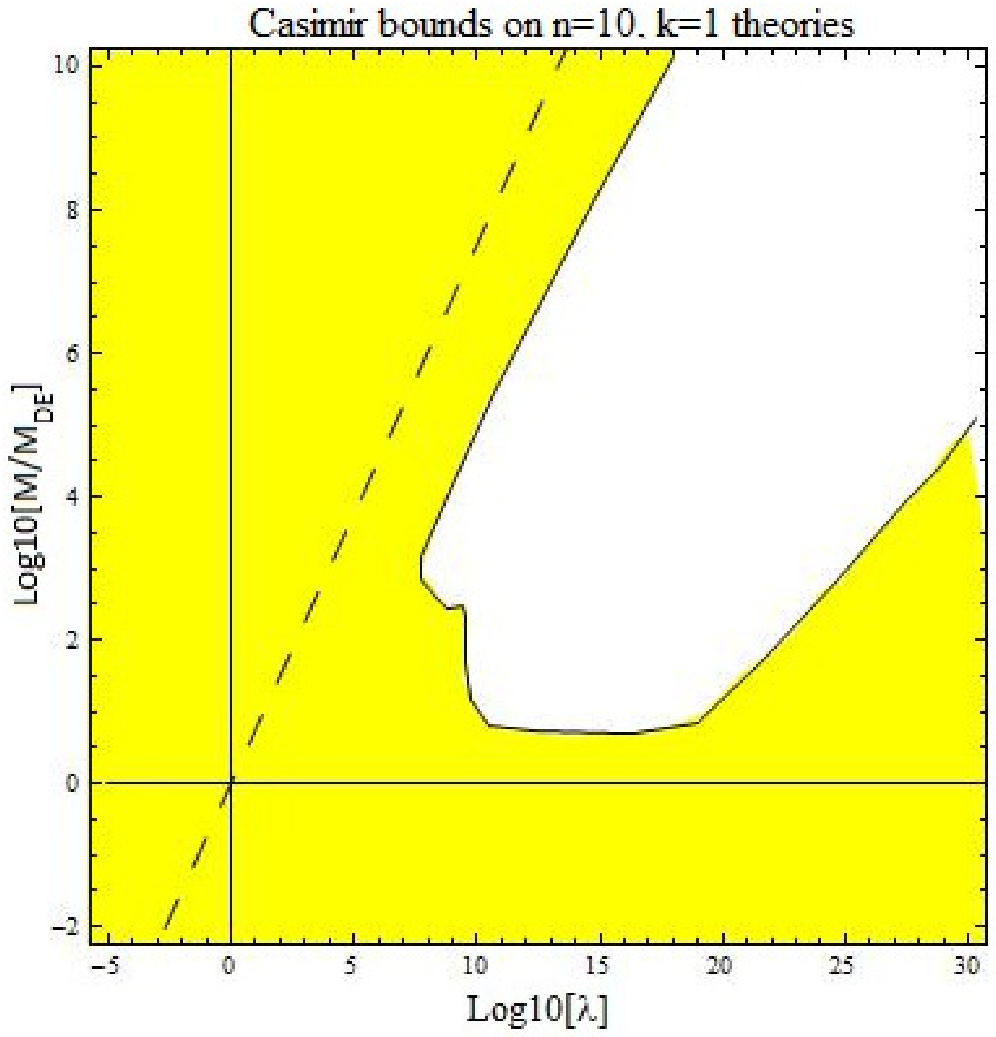}
\caption{Constraints on chameleon theories coming from experimental searches for the Casimir force. The shaded area shows the regions of parameter space that are allowed by the current data. The solid horizontal black lines indicate the cases where $M$ and $\sigma$ take 'natural values'. The solid vertical lines show when $M_{\beta} = H_0$. The dashed black line indicates when $|\beta,_{\phi_c}|M_{\rm pl}=1$ for $\rho_c = \mathcal{O}(1g/cm^3)$. The amount of allowed parameter space increases with $n$.}
\label{casimirbounds}%
\end{figure}

To date, the most accurate measurements of the Casimir force over separations $d = 0.16-1.2\mu m$ have been made by Decca et al. in a series of three experiments taking place between 2003 and 2007 \cite{Decca:2003zz,Decca:2005yk,Decca:2007yb}. We define $P =\frac{F}{A}$ to be the total measured pressure between two parallel plates. Using their most recent experiment, described in Ref. \cite{Decca:2007yb}, Decca et al. found the following 95\% confidence intervals on $\Delta P = P-P_{\rm Casimir}$: at $d = 162nm$, $|\Delta P| < 21.2mPa$, at $d = 400nm$, $|\Delta P| < 0.69mPa$ and at $d = 746nm$, $|\Delta P| < 0.35mPa$. The resulting bounds are shown in Fig.~\ref{casimirbounds}. The area of allowed parameter space grows with $n$ and $k$.
\subsection{Combined bounds}
\begin{figure}%
\centering
\includegraphics[width=0.8\columnwidth]{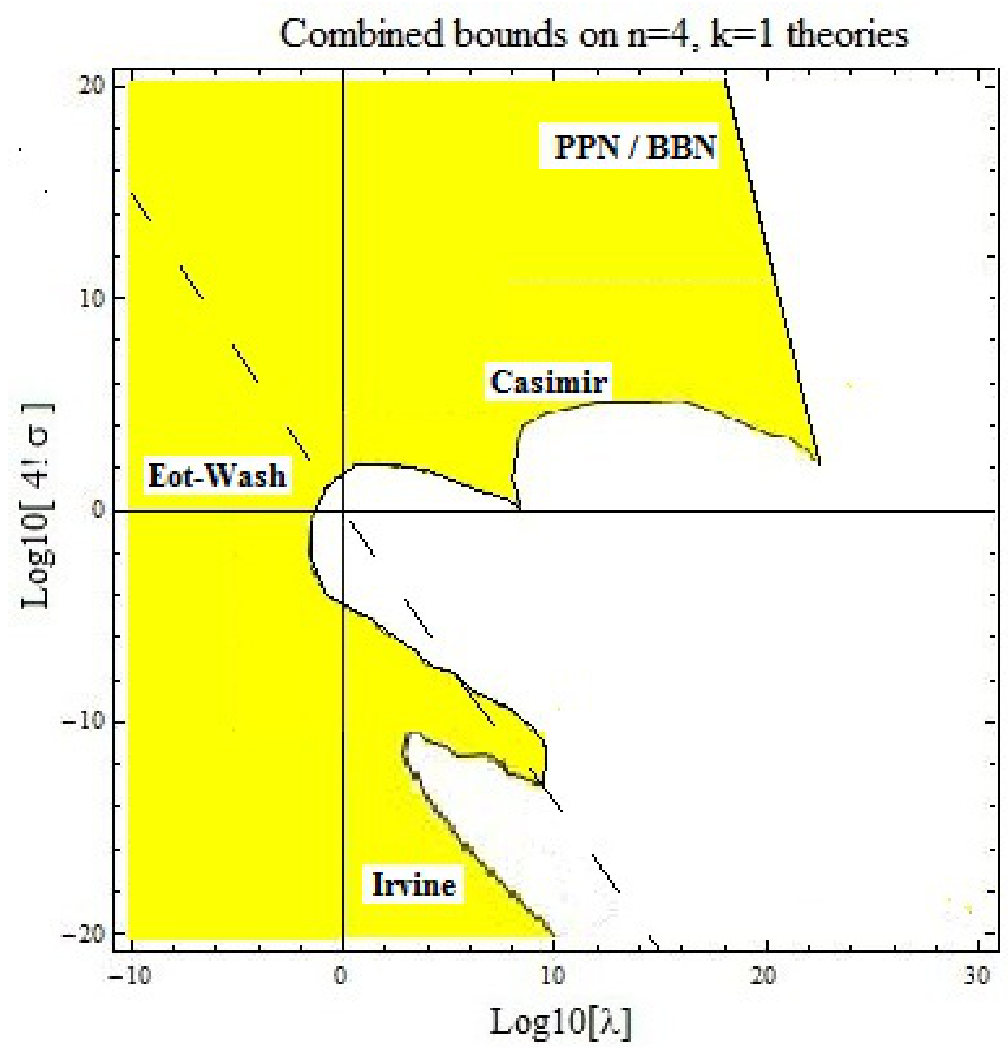}\\
\includegraphics[width=0.8\columnwidth]{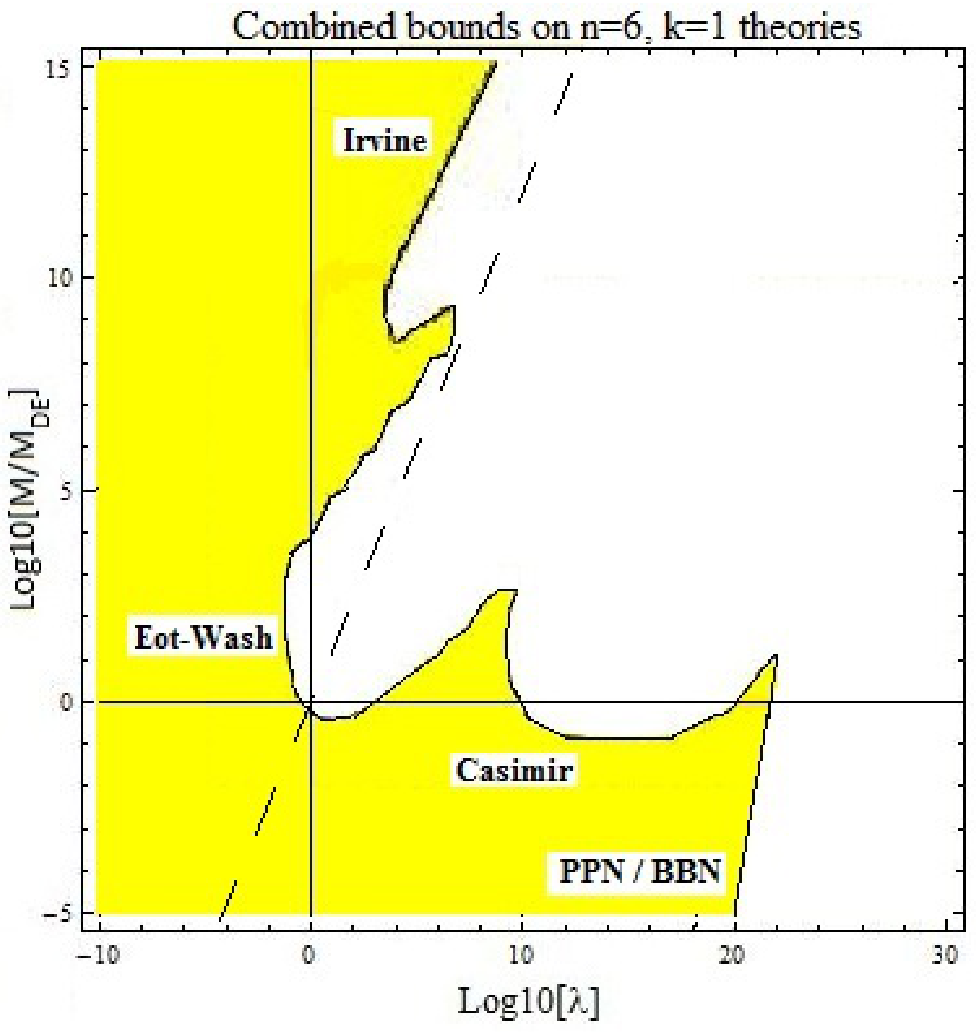}\\
\includegraphics[width=0.8\columnwidth]{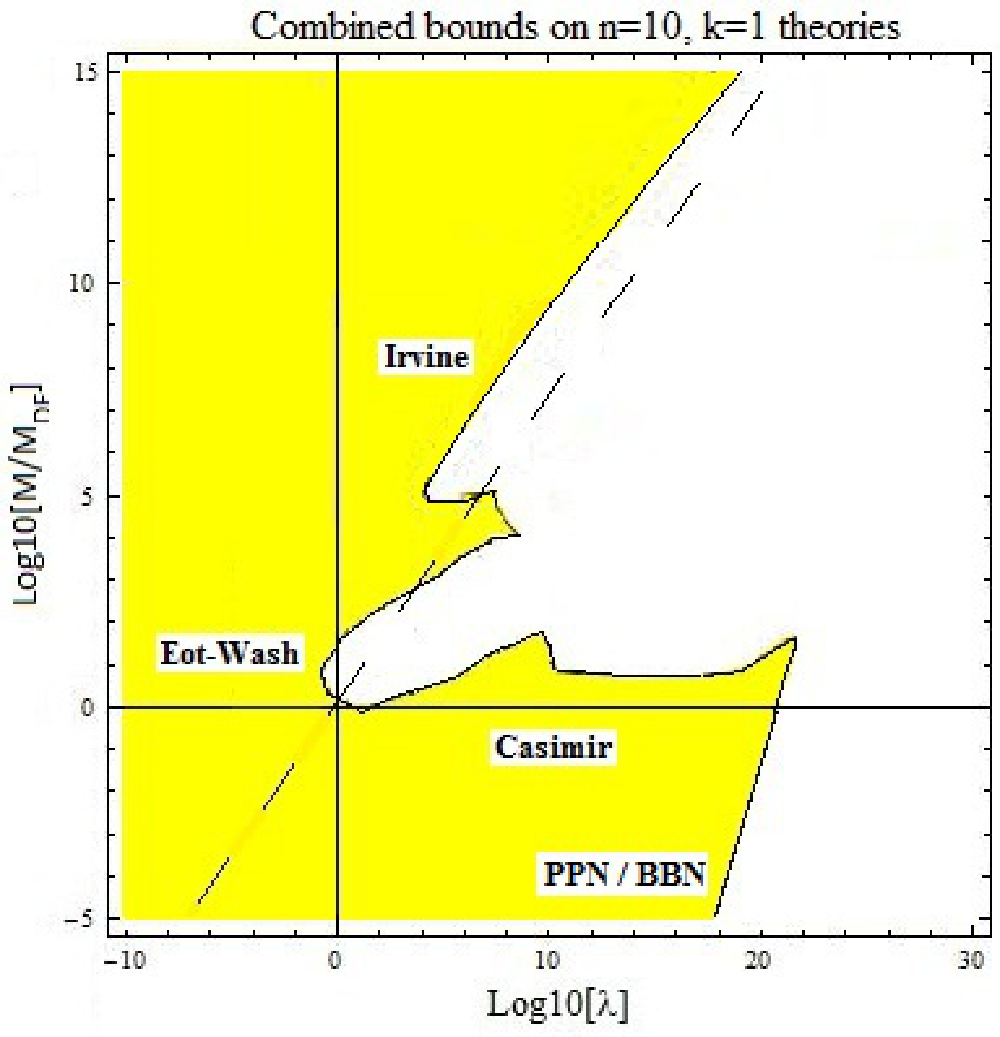}
\caption{Combined constraints on chameleon theories. The shaded area shows the regions of parameter space that are allowed by the current data. The solid horizontal black lines indicate the cases where $M$ and $\sigma$ take 'natural values'. The solid vertical lines show when $M_{\beta} = H_0$. The dashed black line indicates when $|\beta,_{\phi_c}|M_{\rm pl}=1$ for $\rho_c = \mathcal{O}(1g/cm^3)$. The amount of allowed parameter space increases
with $n$.}%
\label{combinedbounds}%
\end{figure}

The chameleon theories considered in this work have a four-dimensional parameter space, spanned either by
$M$ and $\lambda$ ($n>4$), or by $\sigma$ and $\lambda$ ($n = 4$). We combine the constraints found in sections above to bound the
values of $\lambda$ and $M$ (or $\sigma$) for different $n$ taking $k=1$ for simplicity. The constraints for $n = 4,6,10$ with $k=1$ are shown in Fig.~\ref{combinedbounds}. In these figures we have included all the bounds coming from the Eöt-Wash experiment, as well as those coming
from Casimir force searches. We also include the bounds (labeled Irvine) coming from another search for
Yukawa forces. In general, the larger $n$ (and $k$), the larger the region of allowed parameter space. This is the case because, in a fixed density background, the chameleon
mass, $m_c$, scales as $M^{-\frac{(n-4)(2+k)}{n+k}}\sigma^{\frac{2+k}{n+k}}$ and therefore $m_c$ increases with $n$ and $k$ since the exponents are monotonous functions of $n$ and $k$. The larger $m_c$ is, in a given background, the stronger the chameleon mechanism, and a stronger chameleon mechanism tends to lead to looser constraints. The chameleon mechanism also becomes stronger in the limits $M\to 0$ or $\sigma \to \infty$, and all of the constraints are more easily satisfied in these limits. The interesting region of the parameter space is when $M\sim M_{DE}$ and $\lambda\sim 1$. When $\lambda$ is very small, the chameleon mechanism is so weak that, in all cases, the chameleon behaves like a standard (non-chameleon) scalar field and the bounds depends solely on the value of $|\beta,_{\phi_b}|M_{\rm pl}$. It is clear that $\lambda\gg 1$ (which implies $|\beta,_{\phi_c}|M_{\rm pl}\gg 1$) is very much allowed for a large class of chameleon theories. This is in agreement with what was found for the SCM in \cite{Mota:2006fz}.
\section{Conclusions}
We have studied a scalar-tensor theory with a field dependent coupling (assumed to be of the form of an inverse power-law) and a power-law self-interacting potential.

Our main result is that this theory exhibits the chameleon mechanism as found in the original chameleon proposal \cite{Khoury:2003rn}. Thus, the theory presented here is a chameleon field theory and many of the familiar properties of the standard chameleon model carry over to this new setup. The thin-shell solutions we find do not have the geometrical interpretation as found in the SCM. Nevertheless we can define a thin-shell factor which is analogous to the thin-shell factor in the SCM and which describes the suppression of the fifth-force.

The effective coupling in the thin-shell solutions are of the same form as found in the SCM when using a power-law potential, compare (\ref{eq:betaeff}) with \cite[Eq. 24]{Mota:2006fz}. This result is not surprising since the self-interactions are dominating when we have a thin-shell.

If we look at the bounds computed here we see that the natural values $M=M_{DE}$ ($\sigma = \frac{1}{4!}$) together with $\lambda\sim 1$ are ruled out by the Eöt-Wash experiment for $n\lesssim 10$. But we have shown that there exists a large region, $M\lesssim M_{DE}$, in parameter space which is allowed by experiments and in which $|\beta,_{\phi_c}|M_{\rm pl}\gg 1$. These results are equivalent to what has been found in the SCM, and are due to the thin-shell effect.

Assuming that the scalar field plays the role of dark energy, we need to fine-tune the mass-scale in the coupling sector, namely, we have to demand that $M_{\beta}\sim H_0$. It should be noted that even though this mass scale has an unnaturally small value, it is not strictly this scale which determines the coupling strength to gravity: it is given by $|\beta,_{\phi_0}|M_{\rm pl}$ which can be of order unity or larger. If we redefine the field by (\ref{equivalent_action}) then this fine-tuning can be removed and the resulting action has only one fine-tuned mass-scale.

In cosmology the field is well behaved and can act as a dark-energy field causing the late time acceleration of the universe in the same manner as the SCM. The evolution of the density parameters when the field is slow rolling along the attractor is very close to that of $\Lambda$CDM. The cosmological properties of this model are left for future work.

\acknowledgments
CvdB is in part supported by STFC. DFM thanks the Research Council of Norway FRINAT grant 197251/V30. NJN is supported by Deutsche Forschungsgemeinschaft, project TRR33 and program Pessoa 2010/2011. DFM and NJN are also partially supported by project CERN/FP/109381/2009 and PTDC/FIS/102742/2008. 
One of us (Ph.~B.) would like to thank the EU Marie Curie Research \& Training network ``UniverseNet" (MRTN-CT-2006-035863) for support.

\end{document}